\documentclass[final,3p,times,11pt]{elsarticle}

\usepackage{amssymb}

\usepackage{mathtools}
\usepackage{commath}
\usepackage{braket}
\usepackage{array}
\usepackage{xcolor}
\usepackage{multirow}
\usepackage{caption}
\usepackage{subcaption}
\usepackage{todonotes}
\usepackage[mathscr]{eucal}
\usepackage[ruled, vlined]{algorithm2e}
\usepackage{url}
\usepackage{breakurl}
\usepackage[breaklinks]{hyperref}
\usepackage{cuted}
\usepackage{lineno}
\usepackage{xurl}

\newcommand{\cb}{\color{blue}}
\newcommand{\cn}{\color{black}}

\biboptions{sort&compress, square}

\newcounter{bla}

\begin{document}

\begin{frontmatter}



\title{TTDFT: A GPU accelerated Tucker tensor DFT code for large-scale Kohn-Sham DFT calculations}


\author[a]{Chih-Chuen Lin}
\author[a,b]{Vikram Gavini \corref{author}}

\cortext[author] {Corresponding author.\\\textit{E-mail address:} vikramg@umich.eduu}
\address[a]{Department of Mechanical Engineering, University of Michigan, MI 48109-2125, United States}
\address[b]{Department of Materials Science \& Engineering, University of Michigan, MI 48109-2125, United States}

\begin{abstract}
We present the Tucker tensor DFT (TTDFT) code which uses a tensor-structured algorithm with graphic processing unit (GPU) acceleration for conducting ground-state DFT calculations on large-scale systems. The Tucker tensor DFT algorithm uses a localized Tucker tensor basis computed from an additive separable approximation to the Kohn-Sham Hamiltonian. The discrete Kohn-Sham problem is solved using Chebyshev filtering subspace iteration method that relies on matrix-matrix multiplications of a sparse symmetric Hamiltonian matrix and a dense wavefunction matrix, expressed in the localized Tucker tensor basis. These matrix-matrix multiplication  operations, which constitute the most computationally intensive step of the solution procedure, are GPU accelerated providing $\sim$8-fold GPU-CPU speedup for these operations on the largest systems studied. The computational performance of the TTDFT code is presented using benchmark studies on aluminum nano-particles and silicon quantum dots with system sizes ranging up to $\sim7,000$ atoms.

\end{abstract}


\end{frontmatter}



{\bf PROGRAM SUMMARY/NEW VERSION PROGRAM SUMMARY}

\begin{small}
\noindent{\em Program Title:} TTDFT: Tucker tensor density functional theory code                                      \\

\noindent{\em CPC Library link to program files:} (to be added by Technical Editor) \\

\noindent{\em Licensing provisions:} LGPL \\ 

\noindent{\em Programming language:} C/C++                                   \\

\noindent{\em External routines/libraries:}\\ TuckerMPI(https://gitlab.com/tensors/TuckerMPI), \\ 
cuBLAS (https://docs.nvidia.com/cuda/cublas/index.html), \\ cuSparse(https://docs.nvidia.com/cuda/cusparse/index.html), \\
ALGLIB (http://www.alglib.net/), Boost (https://www.boost.org/), \\
BLAS (http://www.netlib.org/blas/), LAPACK (http://www.netlib.org/lapack/), \\
PETSc (https://www.mcs.anl.gov/petsc), SLEPc (http://slepc.upv.es) \\

\noindent{\em Nature of problem:} Real-space Kohn-Sham density functional theory calculations using localized Tucker tensor basis. \\

\noindent{\em Solution method:} We present a real-space Kohn-Sham density functional code based on tensor-structured techniques with GPU acceleration. Tensor-structured techniques are adopted for computing a Tucker tensor basis, representing the eigenfunctions of an additive separable approximation to the Kohn-Sham Hamiltonian. The Tucker tensor basis is further localized using $L_1$ regularization to improve the sparsity of the Kohn-Sham Hamiltonian matrix, and improve the computational efficiency and parallel scalability of the proposed algorithm. The solution to the Kohn-Sham problem in the localized Tucker tensor basis is computed using the Chebyshev filtering subspace iteration (ChFSI) method.\\

\noindent{\em Restrictions:} The code works with Troullier-Martin (TM) pseudopotentials in Kleinman-Bylander form. The current release supports only non-periodic DFT calculations with the local density approximation (LDA) for exchange-correlation functional.\\

\noindent{\em Additional comments:} This TTDFT project uses GitHub via Git, a free distributed version control software. The archived version at the time of submission of this work can be found on the CPC program library through program files DOI provided above. The GitHub repository of this project can be found on \cb \url{https://github.com/ttdftdev/ttdft_public}\cn.\\\

\end{small}

\clearpage

\section{Introduction} \label{ch:introduction}
Electronic structure calculations have provided many insights into the quantum mechanical properties of various materials over the past few decades. Density functional theory (DFT)~\cite{DFT_HK, DFT_KS}, owing to the great balance it provides between accuracy and computational efficiency, has emerged as the workhorse of electronic structure calculations. DFT reduces the Schr\"{o}dinger equation involving the many-body wavefunction in 3$N_{e}$ spatial coordinates ($N_{e}$ denoting the number of electrons), to an equivalent problem of non-interacting electrons in a mean-field that is dependent on the electron density---a variable in only 3 spatial coordinates, thus substantially reducing the computational complexity. While DFT is exact in principle, the many-body quantum mechanical interactions are encapsulated in the exchange-correlation (XC) functional whose form is unknown, and approximate models are used to model the XC functional. The development of increasingly accurate XC functionals is an active area of research~\cite{Becke2014, Jones2015, Mardirossian2017, Perdew2005, Medvedev2017, Kepp2017}.

Despite the wide adoption of DFT for electronic structure calculations, the computational complexity of DFT calculations---conventionally, $O(MN_{e}^{2})$, where $M$ is the number of the basis functions required to achieve desired chemical accuracy, and is usually proportional to the number of electrons in the system ($N_e$)---limits typical DFT calculations to a few hundred atoms. Thus, to improve the computational efficiency of DFT calculations and enable accurate DFT calculations on large-scale systems, it is highly desirable to develop computational methods that can provide systematic convergence and are scalable to large number of MPI tasks, yet with a small basis set. The plane-wave basis, which is the most widely used basis in DFT calculations~\cite{PhysRevB.54.11169,GONZE2002478, QE-2017, Clark2005}, provides systematic convergence, and is well suited for periodic calculations. However, the global nature of the plane-wave basis limits the parallel scalability, and its uniform spatial resolution makes it inefficient for non-periodic systems, such as isolated molecules or clusters. Among the real-space basis sets, the finite-element basis has been demonstrated to be highly scalable~\cite{Motamarri2013b,DFT-FE,SC19Proceedings}---with parallel scalability demonstrated on $\sim200,000$ MPI tasks. However, the number of basis functions required to achieve chemical accuracy is typically much higher than the plane-wave basis. On the other hand, while atomic orbital type basis functions~\cite{MOMethods, QChem, g16, NWChem, FHI-aims} are very efficient---typically involving only few tens of basis functions per atom---systematic convergence is often a concern, especially in metallic systems. Further, the global nature of the basis functions can limit the parallel scalability of calculations.

Recent progress in using tensor-structured techniques for electronic structure calculations has provided a path forward for developing a reduced-order basis that is systematically improvable, efficient, and exhibits good parallel scalability. In particular, an analysis of various molecules has revealed that the electronic structure, in particular the electron density, admits a low-rank Tucker and canonical decomposition~\cite{HACKBUSCH2007697}. Further, \textit{a posteriori} results have shown that the rank required to approximate the electronic density is only weakly dependent on the system size~\cite{BLESGEN20122551}. Based on these observations, a tensor-structured basis was proposed for systematically convergent and efficient large-scale DFT calculations~\cite{Motamarri2016a}. The main ideas included constructing an additive separable approximation of the Kohn-Sham Hamiltonian, and using the eigenbasis of this approximate Hamiltonian---which has a Tucker tensor format---as a reduced-order basis for DFT calculations. Importantly, being the eigenbasis of a Hermitian operator, the resulting Tucker tensor basis provides systematic convergence. Further, being adapted to the Kohn-Sham Hamiltonian, it was demonstrated to be a more efficient basis than the plane-wave basis, requiring fewer basis functions than the plane-wave basis to achieve similar accuracy. However, the global nature of the Tucker tensor basis resulted in a dense Hamiltonian matrix, which limited the accessible system sizes and parallel scalability of the method. In order to alleviate this limitation, we recently proposed an $L_{1}$ localization approach to construct a localized Tucker tensor basis~\cite{Lin2021}, whose span is a close approximation to the subspace spanned by the eigenbasis of the additive separable approximation to the Kohn-Sham Hamiltonian. DFT calculations using the resulting localized Tucker tensor basis were demonstrated on large-scale systems involving many thousands of atoms. Further, this tensor-structured approach was shown to substantially outperform plane-wave implementations even for modest system sizes beyond 2,000 electrons. 

The solution of Kohn-Sham equations in the localized Tucker tensor basis involves many operations that are amenable to acceleration using graphics processing units (GPU). In this work, we present the TTDFT code---Tucker tensor DFT code---that optimizes various parts of the tensor-structured algorithm using GPUs, and provides the code base for conducting large-scale DFT calculations using localized Tucker tensor basis. In particular, we optimize various compute intensive kernels using CUDA library: (i) the matrix-matrix multiplication between the Kohn-Sham Hamiltonian in the localized Tucker tensor basis and the wavefunction matrix expressed in this basis, which appears in the Chebyshev filtering procedure to compute the occupied subspace of the Kohn-Sham Hamiltonian; (ii) the solution of the Kohn-Sham equations by projecting the problem onto the Chebyshev filtered subspace. Our numerical study shows that the implementation substantially accelerates the Chebyshev filtering step---the most time-consuming part in a many-core CPU-based calculation---by $\sim 7\times$ and substantially reduces the wall-times for DFT calculations. Further, we demonstrate the capability of conducting large-scale DFT calculations, with systems as large as $\sim 7,000$ atoms, on GPUs efficiently.     

The remainder of this paper is organized as follows. The Kohn-Sham formulation is presented in Sec.~\ref{ch:kohn-sham dft formulation} for completeness. Section~\ref{ch:numerical simulation} presents the outline of the Tucker tensor algorithm with $L_{1}$ localization for the solution of the Kohn-Sham equations that is implemented in the TTDFT code. We describe the GPU acceleration scheme for improving the computational efficiency in Sec.~\ref{ch:GPU acceleration}. The numerical results from our implementation of GPU accelerated TTDFT code are presented in Sec.~\ref{ch:results}, and we summarize in Sec.~\ref{ch:summary}.

\section{Kohn-Sham DFT formulation} \label{ch:kohn-sham dft formulation}
Kohn-Sham DFT addresses the ground state energy of a quantum mechanical system with $N_a$ atoms and $N_e$ electrons by solving a non-interacting single-particle Schrödinger equation subjected to a mean-field effective potential $v_\mathrm{eff}(\rho; \mathbf{R})$
\begin{linenomath*}\begin{equation}
    \begin{aligned}
    &\mathcal{H} \Psi_{i} = \epsilon_{i} \Psi_{i} \,, \qquad i \in \{1, ..., N_{\mathrm{orb}}\} \\
    &\mathcal{H} = -\frac{1}{2}\nabla^2+v_\mathrm{eff}(\rho; \mathbf{R}) \,.
    \end{aligned}
    \label{eqn:ks eqn} 
\end{equation}\end{linenomath*}
In the above, $\mathcal{H}$ denotes the Kohn-Sham Hamiltonian, $\{\epsilon_{i}, \Psi_{i}\}$ denotes the $i$-th eigenstate, $N_{\mathrm{orb}}$ denotes the number of eigenstates at the lower end of the spectrum that are computed ($N_{\mathrm{orb}}>\frac{N_e}{2}$), and $\mathbf{R}$ denotes the vector with the positions of atoms. 
The electron density---the central quantity of interest in DFT--- is denoted by $\rho=\rho(\mathbf{x})$ in real-space, with coordinates $\mathbf{x}=(x_1, x_2, x_3)$. The electron density is related to the Kohn-Sham orbitals by
\begin{linenomath*}\begin{equation}
    \rho(\mathbf{x}) = 2\sum_{i=1}^{N_{\mathrm{orb}}}f(\epsilon_i; \mu)\left|\Psi_i(\mathbf{x})\right|^2\,,
    \label{eqn:rho_def}
\end{equation}\end{linenomath*}
where $f(\epsilon; \mu)$ denotes the orbital occupancy function, and, in the present work, is represented by the Fermi-Dirac distribution 
\begin{linenomath*}\begin{equation}
    f(\epsilon; \mu) = \frac{1}{1+\mathrm{exp}(\frac{\epsilon - \mu}{k_B T})}\,.
    \label{eqn:frac_occupancy}
\end{equation}\end{linenomath*}
Here, $k_{B}$ is the Boltzmann constant, $T$ is the temperature controlling the smearing of the orbital occupancy function, and $\mu$ is the Fermi energy that is solved using the constraint on the total number of electrons given by
\begin{linenomath*}\begin{equation}
    2 \sum_{i=1}^{N_{\mathrm{orb}}} f(\epsilon_{i}; \mu) = N_{e} \,.
    \label{eqn:fermi_energy}
\end{equation}\end{linenomath*}

The effective potential in the Kohn-Sham Hamiltonian, $v_\mathrm{eff}(\mathbf{\rho})$, is a functional of electron density, and is comprised of three contributions
\begin{linenomath*}\begin{equation}
    v_\mathrm{eff}(\rho) = \frac{\delta E_\mathrm{H}}{\delta \rho} + \frac{\delta E_\mathrm{XC}}{\delta \rho} + v_\mathrm{ext}(\mathbf{x};\mathbf{R})\,.
    \label{eqn:veff_def}
\end{equation}\end{linenomath*}
$E_\mathrm{H}$ is the Hartree energy, which represents the classical Coulomb electrostatic interaction between electrons and is given by (in a non-periodic setting)
\begin{linenomath*}\begin{equation}
    E_\mathrm{H} = \frac{1}{2} \int_{\mathbb{R}^3}\int_{\mathbb{R}^3}\frac{\rho(\mathbf{x}) \rho(\mathbf{x'})}{\abs{\mathbf{x}-\mathbf{x'}}}d\mathbf{x}d\mathbf{x'} = \int_{\mathbb{R}^3} \rho(\mathbf{x}) v_\mathrm{H}(\rho) d\mathbf{x}\,,
    \label{eqn:hartree_energy_def}
\end{equation}\end{linenomath*}
where $v_\mathrm{H}(\rho)$ is the Hartree potential defined by the functional derivative of the Hartree energy
\begin{linenomath*}\begin{equation}
    v_\mathrm{H}(\rho) = \frac{\delta E_\mathrm{H}}{\delta \rho} = \int_{\mathbb{R}^3}\frac{\rho(\mathbf{x'})}{\abs{\mathbf{x} - \mathbf{x'}}} d\mathbf{x'}.
    \label{eqn:hartree_pot_def}
\end{equation}\end{linenomath*}
$E_\mathrm{XC}$ is the exchange-correlation energy, which describes all the many-body quantum mechanical interactions between electrons. The functional derivative of $E_{\mathrm{XC}}$ is labeled as the exchange-correlation potential
\begin{linenomath*}\begin{equation}
    v_\mathrm{XC}(\rho) = \frac{\delta E_\mathrm{XC}}{\delta \rho}\, .
    \label{eqn:pot_xc_def}
\end{equation}\end{linenomath*}
In this work, the local density approximation (LDA) in the form of Ceperley-Alder parametrization with Perdew-Zunger data \cite{LDA_CA, LDA_PZ} is used for the exchange-correlation functional.
The last term in Eq.~(\ref{eqn:veff_def}), $v_\mathrm{ext}(\mathbf{x};\mathbf{R})$, is the electrostatic potential acting on electrons induced by the nuclei. Typically, the core electrons do not participate in chemical reactions, hence a pseudopotential approximation is commonly adopted to replace the all-electron Coulomb potential by a smoother potential acting only on valence electrons. The behavior of the pseudopotential operator $v_{\mathrm{ext}}$ acting on valence electrons is decomposed into a local part $v_{\mathrm{ext}}^{loc}$ and a non-local part $v_{\mathrm{ext}}^{nl}$.  In this work, the norm-conserving Troullier-Martin~\cite{PSP_TM} pseudopotential in Kleinman-Bylander \cite{PSP_KB} form is used. The action of the pseudopotential operator on the Kohn-Sham orbitals in real space is defined as
\begin{linenomath*}\begin{equation}
    v_{\mathrm{ext}}(\mathbf{x};\mathbf{R})\Psi(\mathbf{x}) = v_{\mathrm{ext}}^{loc}(\mathbf{x};\mathbf{R})\Psi(\mathbf{x}) + v_{\mathrm{ext}}^{nl}(\mathbf{x};\mathbf{R})\Psi(\mathbf{x})\,.
    \label{eqn:psp_vnl_vloc}
\end{equation}\end{linenomath*}
\begin{linenomath*}\begin{equation}
    v_{\mathrm{ext}}^{loc}(\mathbf{x};\mathbf{R}) \Psi(\mathbf{x}) = \sum_{J=1}^{N_a}v_{\mathrm{ext}}^{loc, J}(\mathbf{x}-\mathbf{R}_J)\Psi(\mathbf{x})\,,
    \label{eqn:psp_vloc}
\end{equation}\end{linenomath*}
where $v_{\mathrm{ext}}^{loc, J}(\mathbf{x}-\mathbf{R}_J)$ is the corresponding local potential for the $J$-th atom, and $\mathbf{R}_J$ is the coordinate of the $J$-th atom.
\begin{linenomath*}\begin{equation}
    v_{\mathrm{ext}}^{nl}(\mathbf{x};\mathbf{R}) \Psi(\mathbf{x}) = \sum_J^{N_a}\sum_{lm}C^J_{lm}\varphi^J_{lm}(\mathbf{x}-\mathbf{R}_J)\Delta v^J_l (\mathbf{x}-\mathbf{R}_J)\,,
    \label{eqn:psp_vnl}
\end{equation}\end{linenomath*}
where
\begin{equation*}
    C^J_{lm}=\frac{\int \varphi^J_{lm}(\mathbf{x}-\mathbf{R}_J)\Delta v^J_l (\mathbf{x}-\mathbf{R}_J) \Psi(\mathbf{x})d\mathbf{x}}{\int \varphi^J_{lm}(\mathbf{x}-\mathbf{R}_J)\Delta v^J_l (\mathbf{x}-\mathbf{R}_J) \varphi^J_{lm}(\mathbf{x}-\mathbf{R}_J) d\mathbf{x}}    
\end{equation*}
and
\begin{equation*}
    \Delta v^J_l (\mathbf{x}-\mathbf{R}_J) =  v^J_l (\mathbf{x}-\mathbf{R}_J) - v_{\mathrm{ext}}^{loc, J}(\mathbf{x}-\mathbf{R}_J). 
\end{equation*}
Therein, $v^J_l (\mathbf{x})$ is the pseudopotential component of the $J$-th atom corresponding to the $l$ azimuthal quantum number; $\varphi^J_{lm}(\mathbf{x})$ is the single atom pseudo-wavefunction of the $J$-th atom corresponding to the azimuthal and magnetic quantum numbers $l$ and $m$, respectively.

Finally, upon solving Eq.~(\ref{eqn:ks eqn}), Eq.~(\ref{eqn:rho_def}), and Eq.~(\ref{eqn:fermi_energy}) self-consistently in a suitable basis, the ground state energy of the given system can be obtained by
\begin{linenomath*}\begin{equation}
    E_{\mathrm{tot}} = E_{\mathrm{band}} + E_\mathrm{XC} - \int_{\mathbb{R}^3}\rho v_\mathrm{XC} (\rho) d\mathbf{x} - \frac{1}{2}\int_{\mathbb{R}^3} \rho v_\mathrm{H}(\rho) d\mathbf{x} + E_{\mathrm{ZZ}}\,,
    \label{eqn:etotal}
\end{equation}\end{linenomath*}
where 
\begin{equation*}
    E_{\mathrm{band}} = 2\sum_{i=1}^{N_{\mathrm{orb}}} f(\epsilon_i; \mu)\epsilon_i    
\end{equation*}
is the band energy. Finally,  
\begin{equation*}
    E_{\mathrm{ZZ}} = \sum_{I=1}^{N_a} \sum_{J > I}^{N_a} \frac{Z_I Z_J}{\abs{\mathbf{R}_I - \mathbf{R}_J}}
\end{equation*}
is the repulsion energy between nuclei, where $Z_I$ is the valence charge of the $I$-th atom.

\section{Tensor-structured algorithm with $L_{1}$ localization} \label{ch:numerical simulation}
In this section, we present the tensor-structured approach of using $L_{1}$ localized Tucker tensor basis for Kohn-Sham DFT calculations. We note that these ideas have been developed in our prior works~\cite{Motamarri2016a,Lin2021}, and we present the details of the algorithm as implemented in the TTDFT code, before discussing the GPU acceleration strategy for the various compute intensive kernels. In this section, we first provide a brief overview to the Tucker tensor representation, and refer to~\cite{Kolda2009a} for more detailed review. Next, the algorithm to construct localized Tucker tensor basis that is adapted to the Kohn-Sham Hamiltonian is presented. Finally, the solution of the Kohn-Sham equations in the localized Tucker tensor basis by using Chebyshev filtered subspace iteration~\cite{ChFSIoriginal, Zhou2006} is discussed. 

\subsection{Tucker tensor representation} \label{ch:tucker tensor representation}
\begin{figure}[h!]
    \centering
    \includegraphics[width=\textwidth]{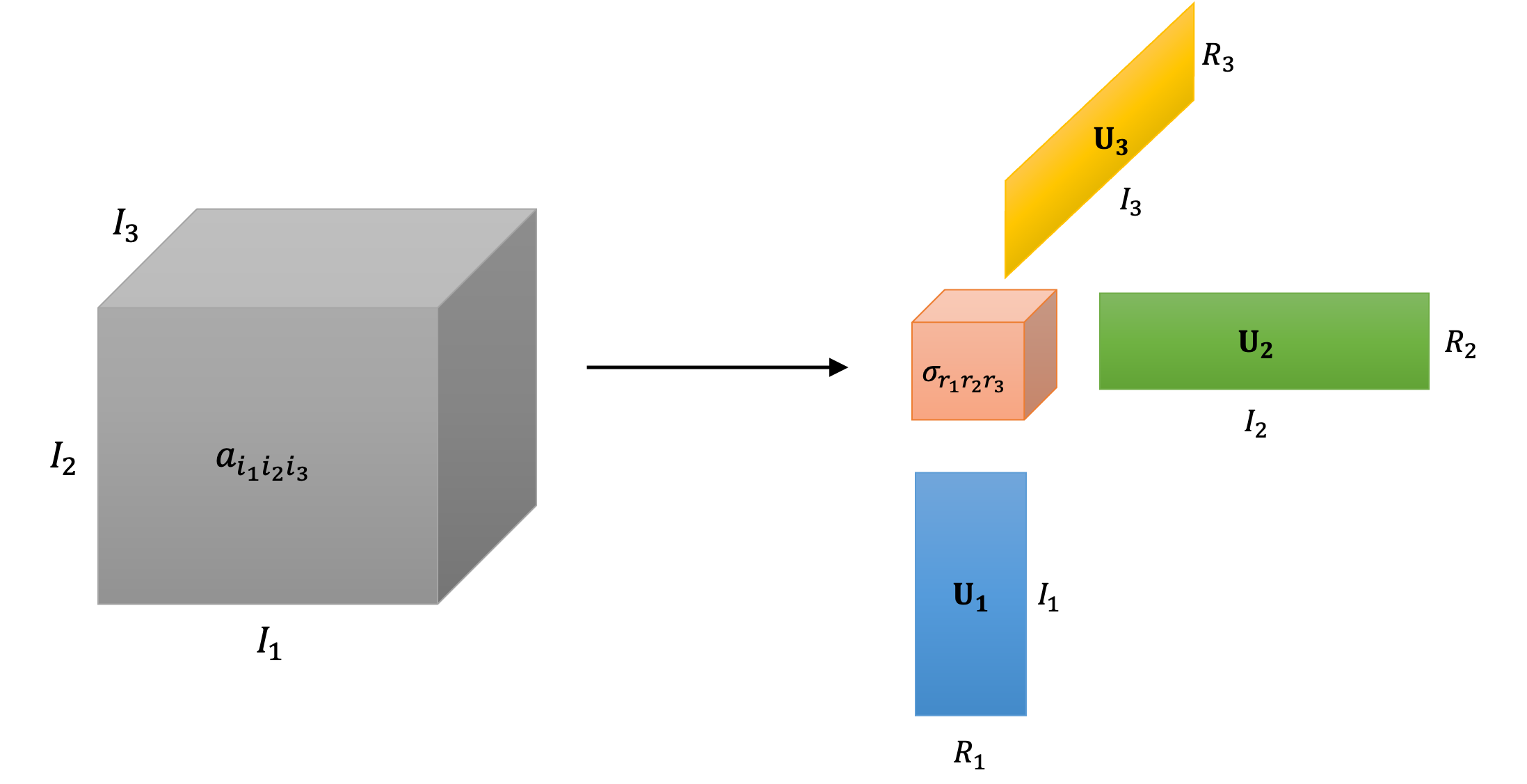}
    \caption{Schematic of Tucker decomposition.}
    \label{fig:tucker_decomp}
\end{figure}

Tucker tensor representation can be regarded as a higher-order generalization of the singular value decomposition of an $N$-dimensional tensor. For an $N$-dimensional tensor, its Tucker tensor representation has the form of a smaller $N$-dimensional tensor and $N$ factor matrices whose column vectors are its rank-1 components. We restrict the discussion to a 3-D tensor as relevant to this work. Let $A \in \mathbb{R}^{I_1 \times I_2 \times I_3}$ be a real-valued 3-D tensor of size $I_1 \times I_2 \times I_3$ indexed by a set of integers $(i_1, i_2, i_3)$ 
\begin{linenomath*}\begin{equation}
    A_{(i_1, i_2, i_3)} = a_{i_1i_2i_3}\,,
    \label{eqn:tensor_entry_def}
\end{equation}\end{linenomath*}
where $i_d \in \{1, 2, ..., I_d\}, I_d \in \mathbb{N}$ and $d \in \{1, 2, 3\}$ denotes the dimensions. A Tucker tensor representation of the tensor $A$ with decomposition rank $\mathbf{R}=(R_1, R_2, R_3)$ has the form
\begin{linenomath*}\begin{equation}
    A \approx A^{(\mathbf{R})} = \sum_{r_1=1}^{R_1} \sum_{r_2=1}^{R_2} \sum_{r_3=1}^{R_3} \sigma_{r_{1} r_{2} r_{3}} \mathbf{u}^
    {r_1}_{1} \mathbf{u}^{r_2}_{2} \mathbf{u}^{r_3}_{3},
    \label{eqn:ttensor_def}
\end{equation}\end{linenomath*}
where $\sigma_{r_{1} r_{2} r_{3}} \in \mathbb{R}^{R_{1} \times R_{2} \times R_{3}}$ denotes the core tensor, $\mathbf{u}_{d}^{r_{d}} \in \mathbb{R}^{I_{d}}$ are the rank-1 components for the factor matrix $\mathbf{U}_{d} \in \mathbb{R}^{I_{d} \times R_{d}}$. A graphical illustration of the Tucker decomposition process is presented in Fig.~\ref{fig:tucker_decomp}. The core tensor could be viewed as the higher-order generalization of singular values and stores the coefficients $\sigma_{r_{1} r_{2} r_{3}}$ for each rank-1 tensor $\mathbf{u}^{r_1}_{1} \otimes \mathbf{u}^{r_2}_{2} \otimes \mathbf{u}^{r_3}_{3}$. The factor matrices can as well be seen as the higher-order correspondence of the matrices comprising the singular vectors.
We note that many approaches have been suggested to perform Tucker decomposition of a given tensor. In this work, we adopt high-order singular value decomposition (HOSVD) techniques for Tucker decomposition, and we refer to \cite{Kolda2009a, Grasedyck2013, Hackbusch} for more details on the various methods for Tucker tensor decomposition. In particular, in this work, we use the TuckerMPI code for performing HOSVD, which is an MPI implementation of tensor operations in Tucker representation. We refer to~\cite{AuBaKo16,BallardKK20} for the library, and details of the implementation.

\subsection{Construction of $L_{1}$ Tucker tensor basis} \label{ch:construction of L1 Tucker tensor basis}
The construction of $L_{1}$ Tucker tensor basis includes the following steps. (i) Compute an additive separable approximation to the Kohn-Sham Hamiltonian in a cuboidal domain $\Omega$ spanned by three 1-D real domains $\omega_{k=1,2,3}$ along the spatial coordinates and enclosing the compact support of the Kohn-Sham wavefunctions~\cite{Motamarri2016a}. The resulting separable approximation to the Kohn-Sham Hamiltonian has the form $\mathcal{H}_{1} (x_1) + \mathcal{H}_{2} (x_2) + \mathcal{H}_{3} (x_3) \approx \mathcal{H} (\mathbf{x})$. We note that the eigenspace of this separable approximation represents a useful reduced-order subspace and the eigenfunctions have a Tucker tensor structure. (ii) While the Tucker tensor basis constructed from the 1-D eigenfunctions of $\mathcal{H}_{k}$ is efficient~\cite{Motamarri2016a}, requiring fewer basis functions to achieve chemical accuracy in comparison to the plane-wave basis, the global nature of this basis limits the computational efficiency and parallel scalability for large-scale DFT calculations. To this end, we employ $L_{1}$ localization, to construct localized 1-D functions that closely approximate the eigen-subspace of $\mathcal{H}_{k}$. (iii) Finally, these localized 1-D functions are used to construct the 3-D localized Tucker tensor basis. In particular, the 3-D Tucker basis is constructed as the tensor product of the localized 1-D functions. We now will elaborate the details of each aspect for constructing the $L_{1}$ localized Tucker tensor basis.

\subsubsection{Separable approximation to the Kohn-Sham Hamiltonian} \label{ch:approximation to KS Hamiltonian}
The separable approximation to the Kohn-Sham Hamiltonian is constructed based on a rank-1 approximation of the eigenfunction corresponding to the lowest eigenvalue~\cite{Motamarri2016a}. To this end, we seek the solution to the lowest eigenstate of Eq.~(\ref{eqn:ks eqn}) to belong to $S = \{ \Psi (\mathbf{x}) | \Psi (\mathbf{x})  = \psi_{x_1} (x_1) \psi_{x_2} (x_2) \psi_{x_3} (x_3) \}$ that has a rank-1 tensor structure. The solution to lowest eigenstate of the Kohn-Sham equation such that it belong to $S$ is equivalent to finding the minimizer of the following energy functional 
\begin{linenomath*}\begin{equation}
    L(\Psi) = \int_{\Omega} \frac{1}{2}\sum_{\ell=1}^{3} \abs{\frac{d\psi_{x_\ell}}{dx_{\ell}}}^2 \prod_ {m \neq \ell}^{3} \psi_{x_m}^2 + \left(v_{\mathrm{eff}}^{loc} (\mathbf{x}) + \lambda \right) \prod_{\ell = 1}^3\psi_{x_\ell}^2
    + \left(\prod_{\ell=1}^3 \psi_{x_\ell}\right)  v_{\mathrm{ext}}^{nl} (\mathbf{x}) \left(\prod_{\ell = 1}^3\psi_{x_\ell}\right)\, d\mathbf{x}\,,
    \label{eqn:1dlagrangian}
\end{equation}\end{linenomath*}
where $\lambda$ denotes the Lagrange multiplier corresponding to normality of the eigenstate, and $v^{loc}_{\mathrm{eff}} = v_{\mathrm{H}} + v_{\mathrm{XC}} + v^{loc}_{\mathrm{ext}}$ denotes the local part of the effective potential.
Upon writing the Euler-Lagrange equations corresponding to variations with respect to $\psi_{x_1}$, $\psi_{x_2}$, $\psi_{x_3}$, and using Tucker tensor decomposition on both $v^{loc}_{\mathrm{eff}}$ and $v^{nl}_{\mathrm{ext}}$, we obtain simultaneous 1-D problems in the form
\begin{linenomath*}\begin{equation}
    \begin{aligned}
        &\mathcal{H}_k \psi_k = -(\lambda + a_k) \psi_k \,,\\
        &\mathcal{H}_k = -\frac{1}{2} \frac{d^2}{dx_k^2} + v^{loc}_k(x_k; \psi_{l \neq k}) + v^{nl}_k(x_k; \psi_{l \neq k}) \,,
    \end{aligned}
    \label{eqn:1dks}
\end{equation}\end{linenomath*}
where $v_k^{loc} (x_k)$, $v_k^{nl} (x_k)$ are the local and the non-local contribution to the 1-D potentials respectively, and $a_{k}$ is a constant parametrized by $\psi_{l \neq k}$. The solution to Eq.~\ref{eqn:1dks} can be obtained via a self-consistent field iteration, and we refer to ~\cite{Lin2021} for complete details on the formulation and solution procedure.

\subsubsection{SOC algorithm and $L_{1}$ localization} \label{ch:SOC algorithm and L1 localization}
The minimizer of Eq.~(\ref{eqn:1dlagrangian}) also yields an additive separable approximation to the Kohn-Sham Hamiltonian ($\mathcal{H}_{1} (x_1) + \mathcal{H}_{2} (x_2) + \mathcal{H}_{3} (x_3))$. We note that the eigenfunctions of this approximate Hamiltonian  are a tensor product of the eigenfunctions of the 1-D Hamiltonians ($\mathcal{H}_k$, $k=1,2,3$), owing to the additive separable structure. Further, these eigenfunctions represent a suitable reduced-order basis for the solution of the Kohn-Sham equations in Eq.~(\ref{eqn:ks eqn}). We note that the plane-wave basis, the mostly widely used basis for DFT calculations, represent the eigenfunctions of the Laplace operator, whereas the eigenfunctions of the additive separable approximation also have some information of the Kohn-Sham potential and are expected to have better approximation properties. In fact numerical studies~\cite{Motamarri2016a} have shown exponential convergence with increasing basis size, and chemical accuracy was attained with fewer basis functions in comparison to plane-wave basis. While efficient in terms of basis size, this basis is spatially extended and results in a dense discrete Kohn-Sham Hamiltonian matrix that limits the computational efficiency and parallel scalability. This limitation was addressed in our recent work~\cite{Lin2021}, where localized 1-D functions are generated such that the span of these functions is a close approximation to the space spanned by the eigenbasis of $\mathcal{H}_k$. In particular, the localized functions are generated using an $L_1$ localization technique by solving the following constraint minimization problem 

\begin{linenomath*}\begin{equation}
     \min_{\boldsymbol{\psi}'_k \in \mathbb{R}^{n \times N_{k}}} \frac{1}{\mu} \abs{\boldsymbol{\psi}'_k} + \mathrm{Tr}({\boldsymbol{\psi}'_k}^T \mathbf{H}_k \boldsymbol{\psi}'_k) \quad \textrm{with}\,\, {\boldsymbol{\psi}'_k}^T\boldsymbol{\psi}'_k=I,
    \label{eqn:L1_constraint}
\end{equation}\end{linenomath*}
where $\mathbf{H}_{k}$ is the 1-D separable approximation to the Kohn-Sham Hamiltonian represented in a suitable orthogonal basis, $\boldsymbol{\psi}'_{k}$ is a matrix comprising of $N_{k}$ trial 1-D functions represented in the orthogonal basis, $n$ denotes the number of rows (and columns) of $\mathbf{H}_{k}$, $N_{k}$ is the number of 1-D functions to be computed.

The splitting orthogonality constraint algorithm (SOC) is used in this work for solving the constraint minimization problem in Eq.~(\ref{eqn:L1_constraint}). We refer to~\cite{Lin2021} for details of the SOC algorithm used in the context of the tensor-structured algorithm for generating 1-D $L_{1}$ localized functions from the separable approximation to the Kohn-Sham Hamiltonian. We also refer to ~\cite{Compressed_mode_osher, SOC} for more information on the method and its wider applications.

\subsubsection{3-D localized tensor-structured basis construction} \label{ch:3-D Tucker tensor basis}
Upon solving the constraint minimization problem in Eq.~(\ref{eqn:L1_constraint}), the localized 1-D functions $\psi_{x_1, r_1}^L$, $\psi_{x_2, r_2}^L$, $\psi_{x_3, r_3}^L$ are computed, and the number of localized 1-D functions in each direction---denoted by $R_1$, $R_2$ and $R_3$---constitutes the Tucker rank in each direction of the 3-D localized Tucker tensor basis. The 3-D Tucker tensor basis is given by the tensor product of the 1-D localized functions as

\begin{linenomath*}\begin{equation}
	T^L_K(\mathbf{x}) = \psi_{x_1, r_1}^L (x_{1}) \psi_{x_2, r_2}^L (x_{2}) \psi_{x_3, r_3}^L (x_{3}),
	\label{eqn:3-D basis}
\end{equation}\end{linenomath*}
where $1 \leq r_d \leq R_d$ ($d=1,2,3$) and $K$ is the composite index $K = (r_1, r_2, r_3)_{1 \leq r_d \leq R_d}$. The space spanned by the 3-D localized tensor-structured basis functions are denoted as $\mathbb{T}^L$.

\subsection{Discrete Kohn-Sham eigenvalue problem} \label{ch:projected Hamiltonian Computation}
The discrete Kohn-Sham Hamiltonian in the localized tensor-structured basis functions $T^L_{I}$ is given by
\begin{linenomath*}\begin{equation}
	H^L_{I, J} = \Bra{T^L_{I}}  -\frac{1}{2}\nabla^2+v_\mathrm{eff}(\rho; \mathbf{R}) \Ket{T^L_{J}},
	\label{KS-Hamiltonian projection}
\end{equation}\end{linenomath*}
where $I$ and $J$ are composite indices $I=(i_1, i_2, i_3)$, $J=(j_1, j_2, j_3)$. We note that in practice, the effective potential $v_{\mathrm{eff}}$ is represented in Tucker format to take advantage of the efficient tensor-structured calculation for computing entries of the Kohn-Sham Hamiltonian. Upon computing the Hamiltonian matrix, the matrix elements that are smaller than a prescribed tolerance are set to zero to attain better sparsity in the discrete Hamiltonian. Further, while the 3-D localized Tucker tensor basis can be computed for every self-consistent field (SCF) iteration of the Kohn-Sham problem, numerical studies have suggested that it suffices to construct the Tucker tensor basis in the first iteration and keep this fixed during the SCF iteration~\cite{Lin2021}, as the error from the separable approximation of the Kohn-Sham Hamiltonian typically dominates the SCF error. Owing to the orthonormality of the 3-D localized Tucker tensor basis, the discrete Kohn-Sham eigenvalue problem is given by

\begin{linenomath*}\begin{equation}
		\mathbf{H}^{L}\mathbf{\Psi} = \mathbf{\Psi}\mathbf{\Lambda} \,.
	\label{eqn:localized standard ks problem}
\end{equation}\end{linenomath*}

\subsection{Chebyshev filtered subspace iterative (ChFSI) method}
The standard eigenvalue problem in Eq.~(\ref{eqn:localized standard ks problem}) is solved using the Chebyshev filtering subspace iteration (ChFSI) method~\cite{Zhou2006}. The ChFSI method has been demonstrated to be an effective method for large-scale real-space DFT calculations~\cite{DFT-FE,SC19Proceedings}. In every SCF iteration, the ChFSI method seeks to compute a good approximation to the subspace spanned by the occupied states of the Kohn-Sham Hamiltonian. This is realized by taking advantage of the property of Chebyshev polynomials that are bounded in the interval $\left[-1, 1\right]$, but grow rapidly outside this interval. To this end, the discrete Kohn-Sham Hamiltonian is scaled and shifted such that the unwanted spectrum maps to $\left[-1, 1\right]$ and the desired spectrum of the occupied and partially occupied states maps to $\left(-\infty, -1\right)$. Thus, the application of a Chebyshev polynomial filter, constructed from the scaled-and-shifted Hamiltonian, on a set of vectors provides a subspace that is a close approximation to the desired occupied eigenspace. The Chebyshev filtered vectors are orthogonalized using Gram-Schmidt orthogonalization procedure, and the Kohn-Sham eigenvalue problem (Eq.(\ref{KS-Hamiltonian projection})) is solved by projecting the problem onto the Chebyshev filtered subspace. 

\begin{algorithm}[h!]
\SetAlgoLined
\KwIn{$\mathbf{H}^{L}$, $\mathbf{X}$, $m$, $\epsilon_{0}$, $\epsilon_{ub}^{w}$, $\epsilon_{ub}^{uw}$}
\KwOut{$\mathbf{\Psi}$, $\mathrm{diag}\left(\mathbf{\Lambda}\right)$}
1. Chebyshev filtering process \\
    \hspace{6mm} Initialize: $e = \frac{1}{2}\left(\epsilon_{ub}^{uw}-\epsilon_{ub}^{w}\right)$;  $c =  \frac{1}{2}\left(\epsilon_{ub}^{uw}+\epsilon_{ub}^{w}\right)$; $\sigma = \frac{e}{\epsilon_{0} - c}$ \\ 
    \hspace{6mm} $\sigma_{1} = \sigma$; $\gamma = \frac{2}{\sigma_{1}}$; $\tilde{\mathbf{X}} = \frac{\sigma_1}{e}\left(\mathbf{H}^{L} \mathbf{X} - c \mathbf{X} \right)$;\\
    \hspace{6mm} \textbf{for} $i = 2:m$ \\
        \hspace{10mm} $\sigma_{2} = \frac{1}{\gamma - \sigma}$; \\
        \hspace{10mm} $\tilde{\mathbf{X}}_{new} = \frac{2\sigma_{2}}{e}\left(\mathbf{H}^{L} \tilde{\mathbf{X}} - c\tilde{\mathbf{X}} \right) - \sigma \sigma_{2} \mathbf{X}$; \\
        \hspace{10mm} $\mathbf{X} = \tilde{\mathbf{X}}$; $\tilde{\mathbf{X}} = \tilde{\mathbf{X}}_{new}$; $\sigma = \sigma_{2}$; \\
    \hspace{6mm} \textbf{end for} \\
2. Orthonormalize the Chebyshev filtered basis functions, and denote by $\mathbf{X}_F = \mathrm{Orth}(\mathbf{X})$ \\
3. Perform subspace projection: $\mathbf{H}^{L}_{F} = \mathbf{X}^{\mathrm{T}}_{F} \mathbf{H}^{L} \mathbf{X}_{F}$ \\
4. Diagonalize $\mathbf{H}^{L}_{F}$ with eigen-decomposition $\mathbf{H}^{L}_{F} \mathbf{Q} = \mathbf{Q} \mathbf{\Lambda}$ \\
5. Rotate the basis $\mathbf{\Psi} = \mathbf{X}\mathbf{Q}$
\caption{ChFSI~\cite{ChFSIoriginal}}
\label{alg:ChFSI}
\end{algorithm}

The ChFSI method is outlined in Algorithm~\ref{alg:ChFSI} for the sake of completeness, and we refer to ~\cite{ChFSIoriginal,Zhou2006} for further information. In the Algorithm~\ref{alg:ChFSI}, $m$ denotes the Chebyshev polynomial degree; $\epsilon_{0}$ and $\epsilon_{ub}^{w}$ are the lower and upper bound of the wanted spectrum, respectively; $\epsilon_{ub}^{uw}$ is the upper bound of the unwanted spectrum; $\mathbf{X}$ is the input wavefunction matrix; $\mathbf{\Psi}$ is the output wavefunction. As suggested in~\cite{ChFSIoriginal}, the lower bound of the wanted spectrum is used to introduce a further scaling to prevent $\mathbf{X}$ from overflowing. In the first SCF iteration, $\mathbf{X}$ is typically set to either a random full-rank matrix or represented by atomic orbitals, and the Chebyshev filtering is performed using higher polynomial degree $m$. In the subsequent iterations, $\mathbf{X}$ is set to be the resultant $\mathbf{\Psi}$ from the previous SCF iteration, which provides a good guess and thus does not need a large $m$. For the various benchmark systems studied in this work, $m$ is chosen to be $10-20$. 

\section{GPU acceleration} \label{ch:GPU acceleration}
In the solution of the Kohn-Sham equations using the localized Tucker tensor basis, the Chebyshev filtering step in each SCF iteration is the most computationally expensive step for even systems comprising of $\sim10,000$ electrons. The main kernel in the Chebyshev filtering is the sparse-dense matrix-matrix multiplication, and a GPU acceleration of this kernel can result in substantial reductions in the wall-times of the DFT calculation.

To this end, the Hamiltonian matrix $\mathbf{H}^{L}$ and the wavefunction matrix $\mathbf{X}$ are partitioned row-wise. We note that this work takes advantage of band-parallelism to reduce the communication costs and improve parallel scalability, where a subset of wavefunctions are assigned to each group of MPI tasks via sub-communicators. Thus, each GPU owns multiple rows of the Hamiltonian matrix and the wavefunction matrix corresponding to the sub-group. The details of the data layout for the Hamiltonian matrix and the wavefunction matrix are elaborated in following sections. We also remark that the sparsity pattern of $\mathbf{H}^{L}$ is such that the matrix has less sparsity around the diagonal, whereas the sparsity increases away from the diagonal. This structure is due to the spatial locality of the $L_1$ localized Tucker tensor basis. We take advantage of this structure to develop an efficient implementation of the matrix-matrix multiplication kernel in the Chebyshev filtering step. Further, we take advantage of the fact that $\mathbf{H}^{L}$ is symmetric to reduce communication costs. 

The remainder of this section will present our implementation of the various aspects of the TTDFT code that have been GPU accelerated, which include: (i) the details of the data layout for the Kohn-Sham Hamiltonian matrix $\mathbf{H}^{L}$ and the wavefunction matrix $\mathbf{X}$, (ii) the algorithm for matrix-matrix multiplication of $\mathbf{H}^{L}\mathbf{X}$ based on GPU, and (iii) applying the matrix-matrix multiplication kernel for the subspace projection $\mathbf{H}^{L}_{F} = \mathbf{X}_F^{\mathrm{T}} \mathbf{H}^{L} \mathbf{X}_F$.

\subsection{Data layout for $\mathbf{H}^{L}$ and $\mathbf{X}$} \label{ch:data layout}
\begin{figure}[h!]
\centering
\includegraphics[width=\linewidth]{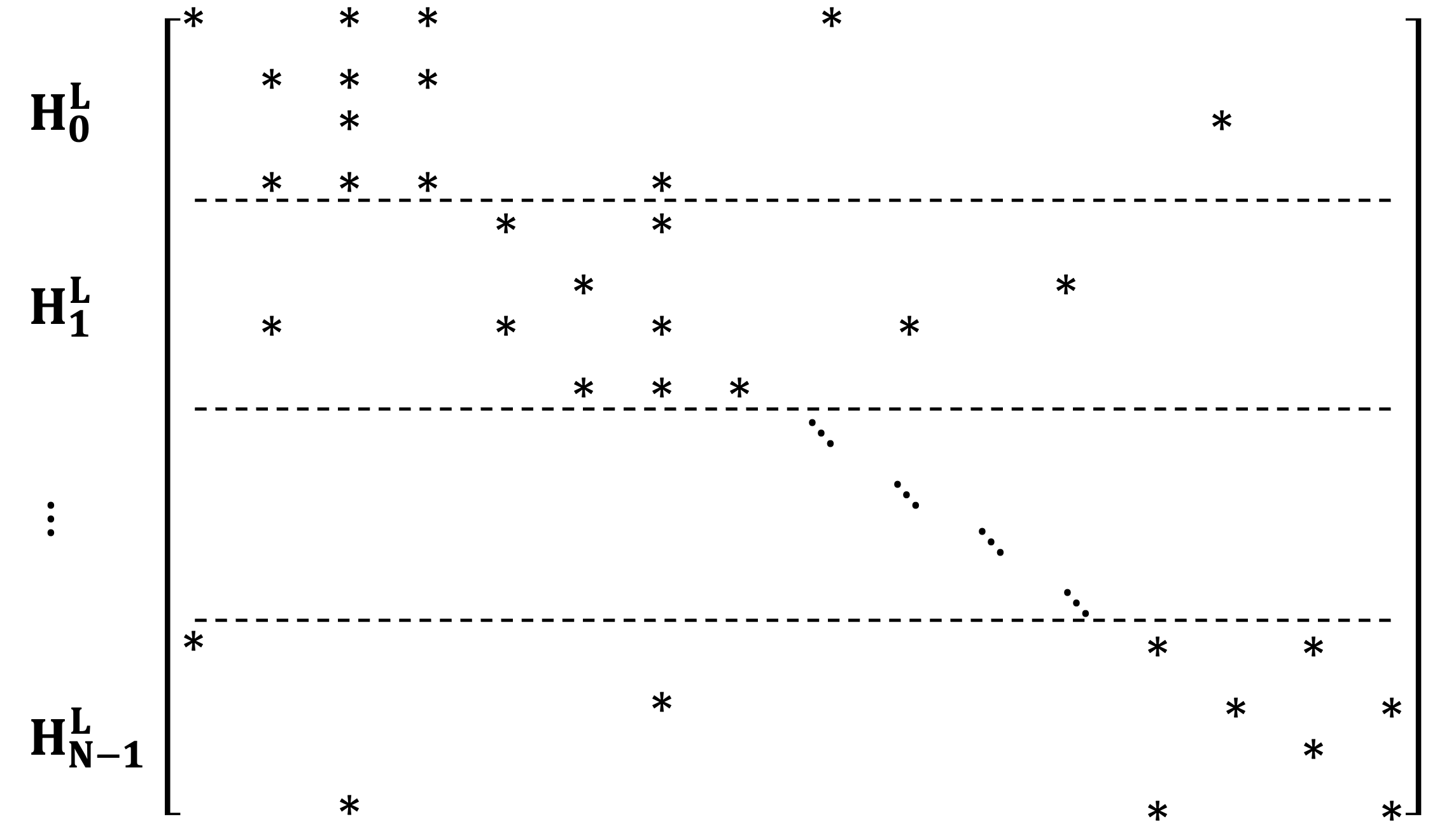}
\caption{Schematic of the distribution of the projected Hamiltonian $\mathbf{H
}^{L}$ on each GPU. The $\mathbf{H}^L_{k}$ ($k=0,1,\ldots N-1$) partition of the projected Kohn-Sham Hamiltonian is assigned to the $k$-th GPU.}
\label{fig:mat scheme}
\end{figure}

Figure~\ref{fig:mat scheme} provides a schematic of the data layout of the sparse Kohn-Sham Hamiltonian matrix $\mathbf{H}^{L}$ in the localized Tucker tensor basis. The ownership of the rows of the Hamiltonian matrix is distributed as evenly as possible so that each GPU shares similar working load. Particularly, given the Hamiltonian matrix $\mathbf{H}^{L}$ of size $M \times M$, the matrix is distributed across $N$ GPUs labeled from $0$ to $N-1$ as shown in Fig.~\ref{fig:mat scheme}. Let $\tau$ be the quotient of $M$ divided by $N$, then block of the Hamiltonian matrix $\mathbf{H}^{L}$ residing on the $k$-th GPU owns the $k \tau$-th to the $((k+1)\tau - 1)$-th rows of the Hamiltonian matrix, and is of size $\tau \times M$. In the case that $M$ is not divisible by $N$, and $\nu$ be the remainder of $M$ divided by $N$, the local block of the Hamiltonian matrix of the first $\nu$ GPUs are adjusted to be of size $(\tau+1) \times M$.

\begin{figure}[h!]
\centering
\includegraphics[width=\linewidth]{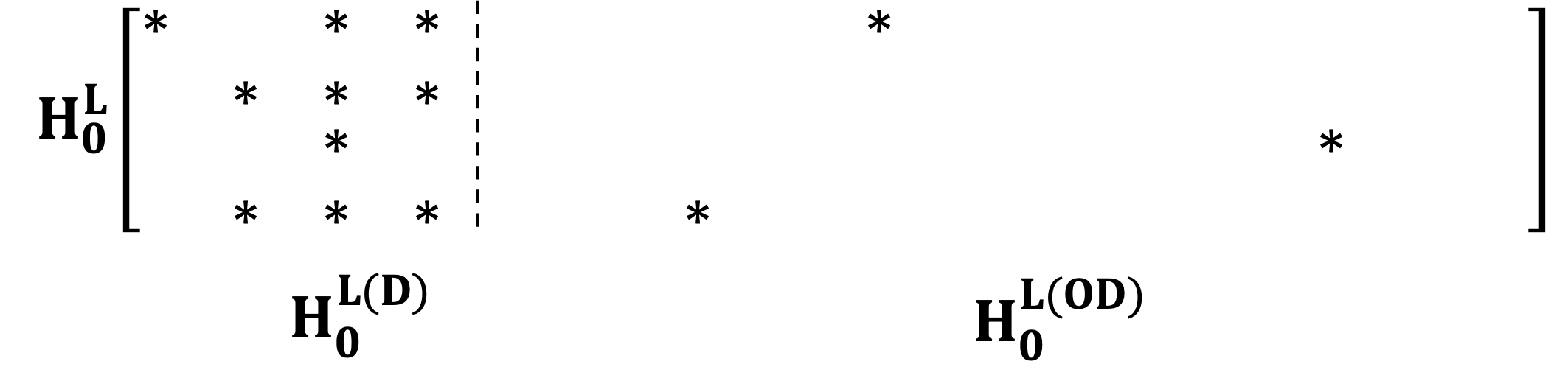}
\caption{Schematic of the data layout of the row-wise partitioned Hamiltonian matrix $\mathbf{H}^{L}$ on the $0$-th GPU.}
\label{fig:local mat scheme}
\end{figure}

We remark that the sparsity pattern of $\mathbf{H}^{L}$ is such that most of the non-zero entries of the matrix are concentrated on and around the diagonal of the matrix, owing to the spatial locality of the $L_1$ localized Tucker tensor basis. Thus, in a tiling of the matrix, the diagonal blocks are much denser compared to the off-diagonal blocks. Thus, the non-zero terms in the diagonal blocks could easily exceed 5\%, which is the suggested minimal sparsity for sparse algorithm to be efficient~\cite{cudasparse}, and deteriorate the overall performance. To this end, the diagonal blocks and the off-diagonal blocks of the Hamiltonian matrix are stored as dense and sparse matrices, respectively. The proposed data layout for the row-wise partitioned matrix on the $0$-th GPU is illustrated in Fig.~\ref{fig:local mat scheme}. The diagonal dense square matrix part of $\mathbf{H}^L_{k}$ is denoted as $\mathbf{H}_{k}^{L(D)}$ and the off-diagonal sparse matrix is denoted as $\mathbf{H}_{k}^{L(OD)}$. The two parts of $\mathbf{H}^L_{k}$ will then be treated using dense and sparse linear algebra library for the matrix-matrix multiplication kernel, respectively. We note that the number of rows owned by each GPU is chosen to be $\sim$30,000 in the current implementation so that the density of the diagonal block of the matrix exceeds 5\%, yet fits in the GPU memory. Above the 5\% threshold, the dense algorithm is generally considered to outperform the sparse algorithm. 

\begin{figure}[h!]
\centering
\includegraphics[width=0.4\linewidth]{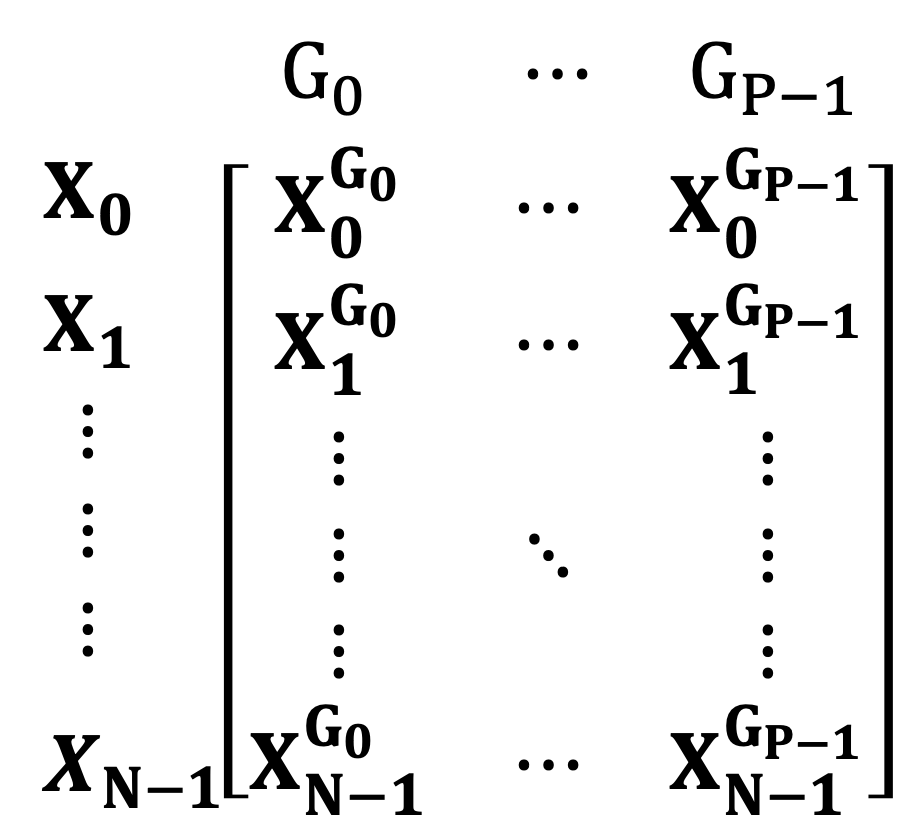}
\caption{Schematic of the data layout of the wavefunction matrix $\mathbf{X}$.}
\label{fig:schematic X}
\end{figure}

The wavefunction matrix $\mathbf{X}$ is of size $M \times N_{\mathrm{orb}}$, where $N_{\mathrm{orb}}$ is the number of computed Kohn-Sham orbitals. Owing to double occupancy of the orbitals for spin-independent Hamiltonian, $N_{\mathrm{orb}}$ is usually chosen to be slightly larger than $N_e/2$, typically $\sim$10-15\% larger. The rows of the wavefunctions are distributed consistently with the row-ownership of the Hamiltonian matrix $\mathbf{H}^{L}$. We note that during the computation of $\mathbf{X}' = \mathbf{H}^{L} \mathbf{X}$, regardless of the implementation, collective communication over either $\mathbf{X}$ or $\mathbf{X}'$ will be needed. The cost for the collective communication is proportional to the number of processors and the data to be communicated within the (sub-)communicator~\cite{mpi_collective, using-mpi}. In a GPU calculation, this also requires data to be transferred from the device memory to the host memory and communication to other processors. Thus, this step will substantially increase the communication cost and deteriorate the overall performance. It is thus desirable to reduce this communication cost. To this end, in addition to the row-wise parallelization, columns of the wavefunction matrix $\mathbf{X}$ are further partitioned into groups (bands) labeled as $G_{p=0...P-1}$, and this is referred to as band parallelization henceforth in keeping with the nomenclature of DFT literature. In the present implementation, the Hamiltonian matrix $\mathbf{H}^L$ is stored on each GPU group. Hence, each group will perform the matrix-matrix multiplication corresponding to the band of wavefunctions, and the collective communication after the matrix-matrix multiplication is only within the processors in the group. The number of processors to be communicated will thus be reduced by a factor $P$ by using band-parallelism. Thus, the communication burden is significantly alleviated, and the overall performance of the Chebyshev filtering step is improved. A schematic illustration of the data layout for the wavefunction matrix $\mathbf{X}$ is provided in Fig.~\ref{fig:schematic X}, where $\mathbf{X}_{k}$ is the portion of the wavefunction matrix having the same row-ownership of the Hamiltonian matrix $\mathbf{H}^{L}_{k}$ in Fig.~\ref{fig:mat scheme}. $\mathbf{X}^{G_{i}}_{k}$ is the portion of the wavefunction matrix $\mathbf{X}_{k}$ belonging to the $G_{i}$ processor group. The data layout for the Hamiltonian matrix $\mathbf{H}^{L}$ and the wavefunction matrix $\mathbf{X}$ are then used to implement the sparse-dense matrix-matrix multiplication kernel, which is subsequently discussed.

\subsection{$\mathbf{H}^{L} \times \mathbf{X}$ implementation} \label{ch:HX implementation}
As noted previously, the Hamiltonian matrix is distributed row-wisely across GPUs in each group $G_{i}$. In the matrix-matrix multiplication, each $\mathbf{H}^{L}_{k} \times \mathbf{X}^{G_i}$, where $\mathbf{X}^{G_i}$ of size $M \times \frac{N_{\mathrm{orb}}}{P}$ is the collection of all $\mathbf{X}^{G_{i}}_{k}$ ($k=0,1,\ldots, N-1$), yields $(\mathbf{H}^{L} \mathbf{X})^{G_{i}}_{k}$ on the $k$-th GPU in the group $G_{i}$. The evaluation of each $(\mathbf{H}^{L} \mathbf{X})^{G_{i}}_{k}$ requires communication of the off-diagonal block of the Hamiltonian matrix $\mathbf{H}^{L(OD)}_{k}$ to all processors other than $k$, as well as collecting information back from those processors. This communication also involves data transfer between the host and the device memory of GPUs, and can severely diminish the performance of the sparse-dense matrix-matrix multiplication kernel.

\begin{figure*}[h!]
\centering
\includegraphics[width=0.8\textwidth]{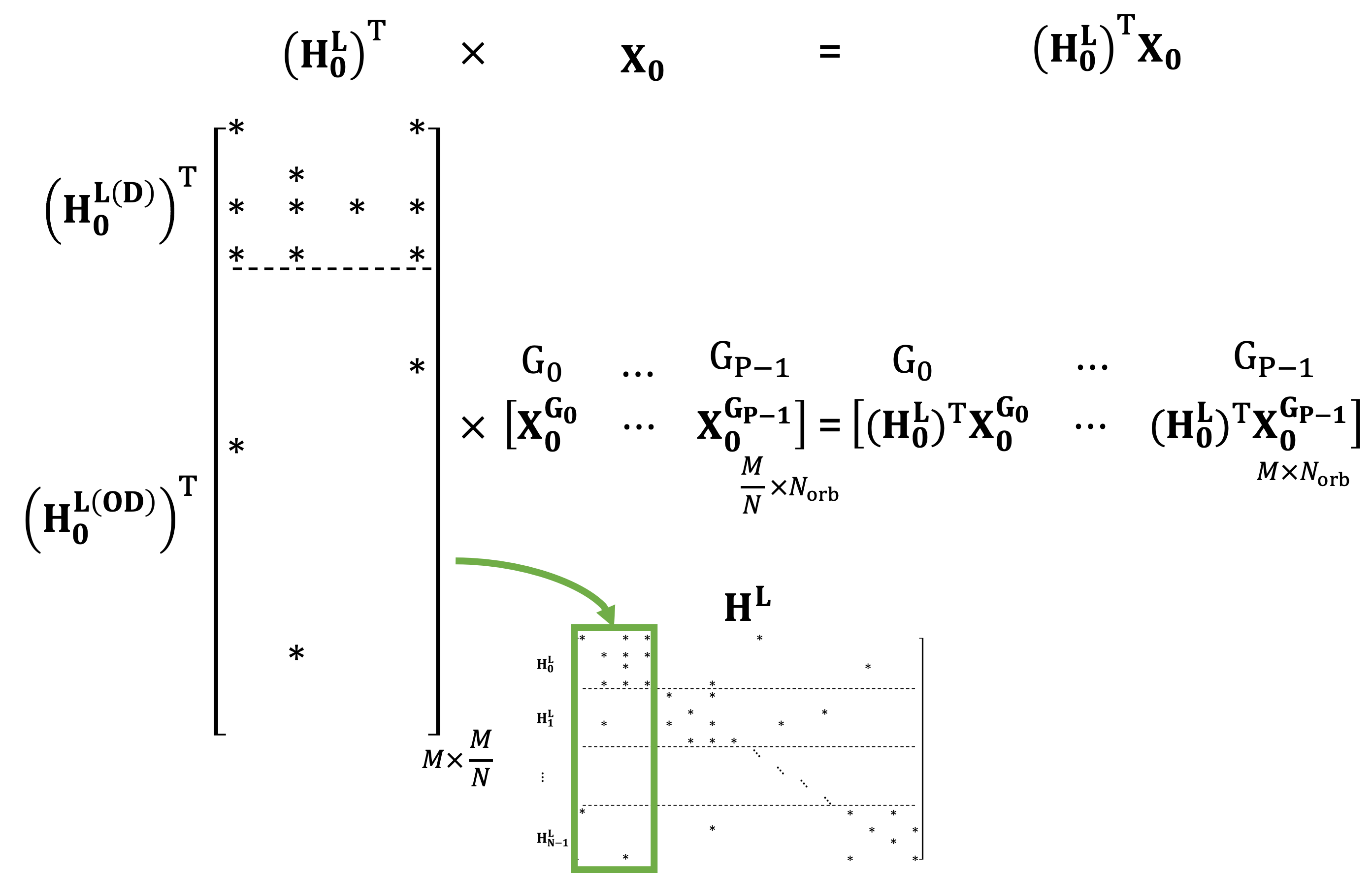}
\caption{Schematic of the computation on $0$-th processor in the evaluation of $\mathbf{H}^{L} \mathbf{X}$. As $\mathbf{H}^{L}$ is symmetric, the transpose of $\mathbf{H}^{L}_{0}$ is equivalent to the columns of $\mathbf{H}^{L}$ enclosed by the highligted box. $(\mathbf{H}^{L(D)}_{0})^{\mathrm{T}}$ and $(\mathbf{H}^{L(OD)}_{0})^{\mathrm{T}}$ denote the dense diagonal square block and the sparse off-diagonal part of the $(\mathbf{H}^{L}_{0})^{\mathrm{T}}$ matrix, respectively. 
}
\label{fig:schematic H*X}
\end{figure*}

In order to avoid the aforementioned communication of $\mathbf{H}^{L(OD)}_{k}$, the matrix-matrix multiplication kernel is recast by taking advantage of the symmetric nature of $\mathbf{H}^{L}$. We note that as $\mathbf{H}^{L}$ is real and symmetric, 
\begin{linenomath*}\begin{equation}
    \mathbf{H}^{L}_{(:, a:b)} = \left(\mathbf{H}^{L}_{(a:b, :)}\right)^{\mathrm{T}}.
    \label{eqn:transpose_equal}
\end{equation}\end{linenomath*}
Eq.~(\ref{eqn:transpose_equal}) states that a column block of the Hamiltonian matrix $\mathbf{H}^{L}_{(:, a:b)}$, which is a matrix containing the $a$-th to $b$-th columns of $\mathbf{H}^{L}$, is equivalent to the transpose of a row block $\mathbf{H}^{L}_{(a:b, :)}$ comprising the $a$-th to $b$-th rows of $\mathbf{H}^{L}$. Further, we note that the evaluation of a matrix-matrix product $\mathbf{C} = \mathbf{A}\mathbf{B}$, where $A$ and $B$ are $m\times n$ and $n\times m$ matrices, is given by $c_{ij} = \sum_{k=1}^{n} a_{ik}b_{kj}$ with $a_{ij}$, $b_{ij}$ and $c_{ij}$ denoting the matrix elements of $\mathbf{A}$, $\mathbf{B}$ and $\mathbf{C}$, respectively. The expression $\sum_{k=1}^{n} a_{ik}b_{kj}$ can also be viewed as a summation over $k$ of the outer product of the $k$-th column vector of $\mathbf{A}$ with the $k$-th row vector of $\mathbf{B}$. Thus, using this interpretation of the matrix-matrix multiplication as a sum of the outer product of column and row vectors of the constituent matrices, we can write
\begin{linenomath*}\begin{equation}
\begin{aligned}
    \mathbf{H}^{L} \mathbf{X} &= \sum_{k=0}^{N-1} \sum_{\xi = a \coloneqq k \tau}^{b \coloneqq (k+1) \tau -1} \mathbf{H}^{L}_{(:,\xi)} \mathbf{X}_{(\xi,:)} 
    := \sum_{k=0}^{N-1} \mathbf{H}^{L}_{(:, a:b)} \mathbf{X}_{(a:b,:)} = \sum_{k=0}^{N-1}  \left(\mathbf{H}^{L}_{(a:b, :)}\right)^{\mathrm{T}} \mathbf{X}_{(a:b,:)} \\
    &= \sum_{k=0}^{N-1} \left(\mathbf{H}^{L}_{k}\right)^{\mathrm{T}} \mathbf{X}_{k} \,.
\end{aligned}
\label{eqn:reformat HX}
\end{equation}\end{linenomath*}
In the above, $\tau = \frac{M}{N}$ follows the definition in Sec.~\ref{ch:data layout}, $\mathbf{H}^{L}_{k}$ and $\mathbf{X}_{k}$ follow the notation in Fig.~\ref{fig:mat scheme} and Fig.~\ref{fig:schematic X}. A schematic for $\left(\mathbf{H}^{L}_{k}\right)^{\mathrm{T}} \mathbf{X}_{k} $ on the $0$-th GPU is illustrated in Fig.~\ref{fig:schematic H*X}. As shown in the figure, the multiplication involves $(\mathbf{H}^{L}_{0})^{\mathrm{T}}$ of size $M \times \tau$ and $\mathbf{X}^{G_{i}}_{0}$ of size $\tau \times \frac{N_{\mathrm{orb}}}{P}$, resulting in matrix $(\mathbf{H}^{L}_{0})^{\mathrm{T}} \mathbf{X}^{G_{i}}_{0}$ of size $M \times \frac{N_{\mathrm{orb}}}{P}$. The final outcome $\mathbf{H}^{L} \mathbf{X}^{G_i}$ can then be obtained by summing over $k$ using $\texttt{Allreduce}$ communication with MPI, as evident from the last equality of Eq.~(\ref{eqn:reformat HX}). We note that both $(\mathbf{H}^{L}_{k})^{\mathrm{T}}$ and $\mathbf{X}_{k}$ are locally stored on the $k$-th GPU. Thus, for each matrix-matrix multiplication call during the Chebyshev filtering step, this approach avoids the MPI communications and overheads associated with transferring data between the device and the host memory for the off-diagonal block of the  Hamiltonian matrix. 

To understand the improvement in the efficiency by avoiding communicating the off-diagonal block of the Hamiltonian matrix, we present a performance comparison by computing $\mathbf{H}^L \mathbf{X}$ using the proposed algorithm and using the method with off-diagonal block communication (henceforth referred to as the general method). On the $i$-th processor, the general method is implemented by sending the off-diagonal blocks of the Hamiltonian matrix to all $j$-th ($j \neq i$) processors whose row-ownership coincide with the columns of the off-diagonal blocks. The off-diagonal block is then multiplied by the locally owned block of wavefunction matrix on the $j$-th processor and the result is reduced back to the $i$-th processor. To ensure the representability of this comparison, we choose the Hamiltonian matrix of $\mathrm{Al}_{147}$, the benchmark system used for performance analysis in the later sections, to run this calculation. This benchmark calculation is run on the GreatLakes HPC cluster with each node comprising 2 Intel Xeon Gold 6148 CPUs with 40 physical cores per node. In this numerical experiment, the general method takes 481.92 cpu-secs and our proposed approach takes 232.18 cpu-secs. The $\sim2\times$ improvement resulting from the communication efficiency, validates the use of the proposed approach.

Next, we turn our attention to leveraging the sparsity structure of $\mathbf{H}^L$ to further optimize the matrix-matrix multiplication kernel. As we noted earlier, the density of the diagonal square block $\mathbf{H}^{L(D)}_{k}$ can be large making sparse linear algebra operations inefficient~\cite{cudasparse}. To this end, we use different linear algebra libraries to treat the dense and the sparse blocks of the Hamiltonian matrix separately. As shown in Fig.~\ref{fig:local mat scheme}, the diagonal blocks $\mathbf{H}^{L(D)}_{k}$ are stored as a dense matrix and the off-diagonal blocks $\mathbf{H}^{L(OD)}$ are stored as a sparse matrix. Further, the matrix-matrix multiplication operation depicted in Fig.~\ref{fig:schematic H*X} is further split into a dense-dense multiplication for the diagonal block \big($\left(\mathbf{H}^{L(D)}_{k}\right)^{\mathrm{T}} \mathbf{X}^{G_{i}}_{k}$\big) and a sparse-dense multiplication for the off-diagonal block \big($\left(\mathbf{H}^{L(OD)}_{k}\right)^{\mathrm{T}} \mathbf{X}^{G_{i}}_{k}$\big). The dense-dense multiplication kernel is implemented using \texttt{cublasDgemm} provided by \texttt{cuBLAS}~\cite{cudablas}, NVIDIA GPU-accelerated implementation for basic linear algebra subroutines (BLAS). On the other hand, the sparse-dense multiplication kernel is computed with \texttt{cusparseDcsrmm} provided by \texttt{cuSPARSE}~\cite{cudasparse}, NVIDIA GPU-accelerated implementation for sparse basic linear algebra subroutines. The two libraries are available in \texttt{CUDA Toolkit} or NVIDIA High performance computing software development kit (\texttt{NVIDA HPC SDK}). Once the computation is completed for the dense-dense and the sparse-dense matrix multiplication, the resultant matrices are assembled as $\left(\mathbf{H}^{L}_{k}\right)^{\mathrm{T}} \mathbf{X}^{G_{i}}_{k}$ (see Fig.~\ref{fig:schematic H*X} for a schematic plot for the $0$-th processor). The assembled matrix $\left(\mathbf{H}^{L}_{k}\right)^{\mathrm{T}} \mathbf{X}^{G_{i}}_{k}$ is transferred back to the host memory. On the host memory, summation over $k$ in Eq.~(\ref{eqn:reformat HX}) is completed using \texttt{MPI\_Allreduce} within the wavefunction group. 

\subsection{Subspace projection: Evaluation of $\mathbf{H}^{L}_{F}$} \label{ch:XtHX}
In the Chebyshev filtering subspace iteration algorithm, upon computing the Chebyshev filtered vectors that represent a close approximation to the eigen-subspace of interest, the Kohn-Sham eigenvalue problem is projected onto the Chebyshev filtered subspace to solve the eigenvalue problem in this subspace. This entails the evaluation of $\mathbf{H}^{L}_{F} = \mathbf{X}^{\mathrm{T}}_F \mathbf{H}^{L} \mathbf{X}_F$ (step 3 in Algorithm~\ref{alg:ChFSI}), where $\mathbf{X}_F$ is comprised of the orthonormalized Chebyshev filtered vectors. The evaluation of $\mathbf{H}^{L}_{F}$ includes a  matrix-matrix multiplication between $ \mathbf{H}^{L}$ and $\mathbf{X}_F$. Thus, it is natural to adopt the strategy discussed in Sec.~\ref{ch:HX implementation} in evaluating $\mathbf{H}^{L} \mathbf{X}_F$. Upon evaluating $\mathbf{H}^{L}\mathbf{X}_F$, this matrix is transferred back to the host memory and left-multiplied with $\mathbf{X}^{\mathrm{T}}$ using MPI-based matrix-matrix multiplication kernel from \texttt{PETSc} library ~\cite{petsc-efficient, petsc-user-ref, petsc-web-page}.

\section{Results} \label{ch:results}
The systematic convergence, accuracy, and efficacy of the Tucker tensor basis and the tensor-structured algorithm for DFT calculations have been established in prior works~\cite{Motamarri2016a,Lin2021}. In particular, it was demonstrated that the Tucker tensor basis was systematically improvable and the basis discretization error decreased exponentially with increasing Tucker rank~\cite{Motamarri2016a,Lin2021}, thus providing spectral convergence similar to plane-wave discretization. We refer to~\cite{Lin2021} for a comprehensive numerical study of the approximation properties of the localized Tucker tensor basis in DFT calculations. Further, a comparative study of the computational efficiency of the localized Tucker tensor basis with a plane-wave basis has revealed that the Tucker tensor basis is not only more efficient in terms of the number of basis functions required to achieve chemical accuracy, but also provides significant computational savings owing to the reduced-order scaling with system size. Benchmark calculations on both systems with and without a gap have revealed that the solution of the DFT problem in the Tucker tensor basis is substantially more efficient than the plane-wave basis for systems beyond 2,000 electrons, with up to $8\times$ improvement in computational efficiency (measured in node-hrs) over plane-wave calculations conducted using Quantum Espresso (cf.~\cite{Lin2021}).

\begin{figure}[htbp]
     \centering
     \begin{subfigure}[b]{0.3\linewidth}
         \centering
         \includegraphics[width=\linewidth]{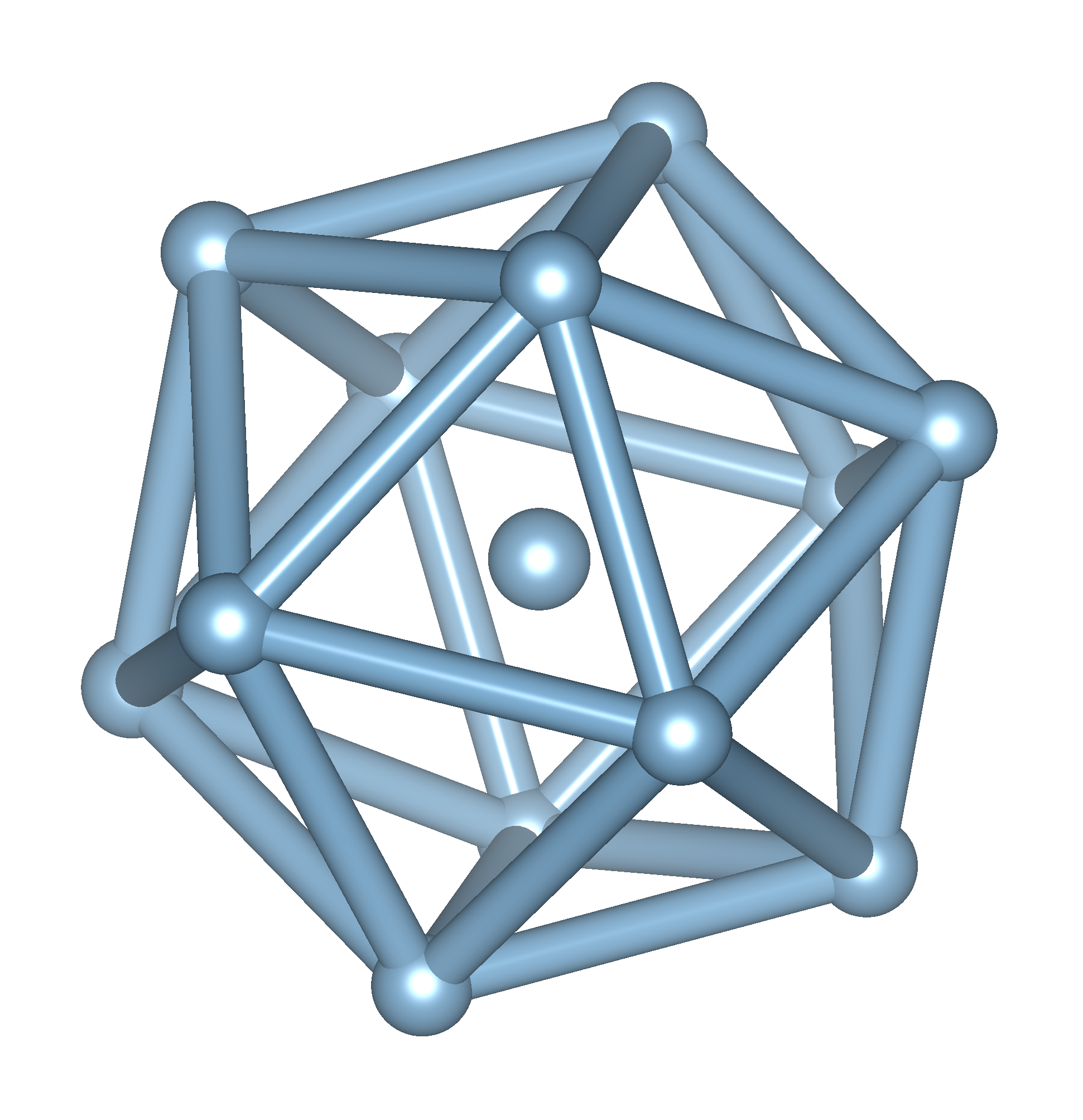}
         \caption{$\mathrm{Al}_{13}$}
         \label{fig:shell1}
     \end{subfigure}
     \hfill
     \begin{subfigure}[b]{0.4\linewidth}
         \centering
         \includegraphics[width=\linewidth]{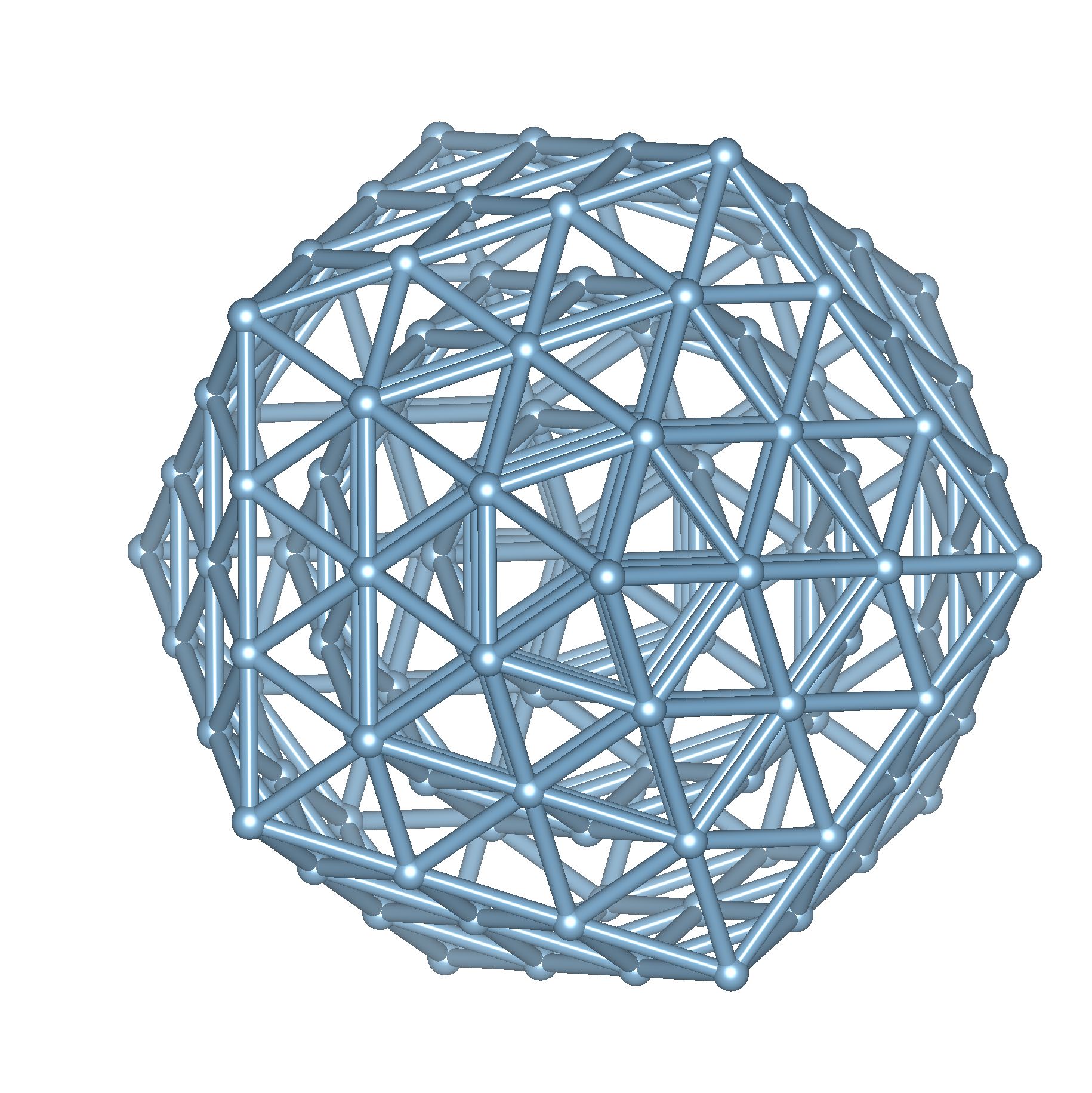}
         \caption{$\mathrm{Al}_{147}$}
         \label{fig:shell3}
     \end{subfigure}
     \caption{Schematics of the benchmark aluminum nano-particles. 
     }
     \label{fig:Al bns model}
\end{figure}

\begin{figure}[htbp]
     \centering
     \begin{subfigure}[b]{0.3\linewidth}
         \centering
         \includegraphics[width=\linewidth]{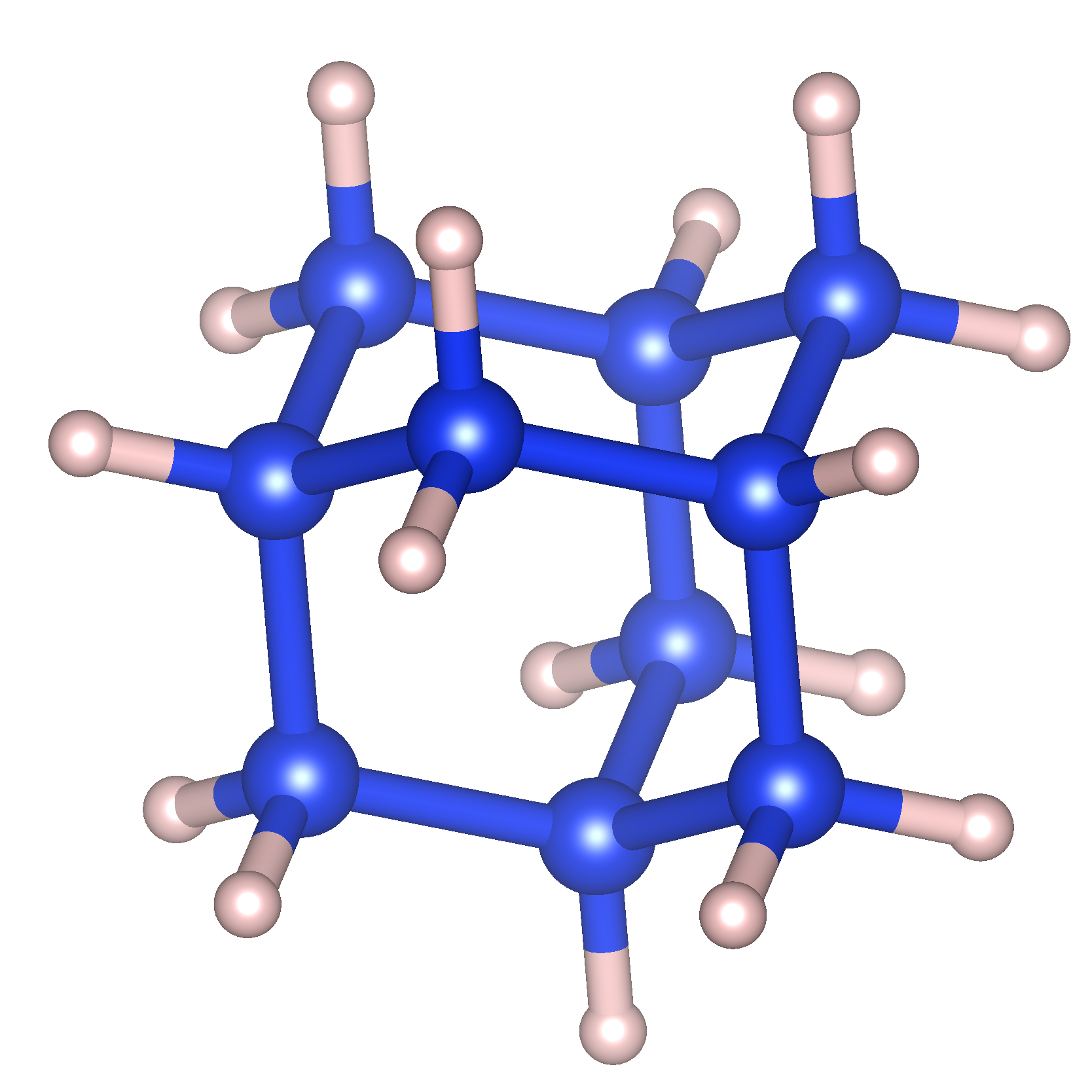}
         \caption{ $\mathrm{Si}_{10}\mathrm{H}_{16}$}
         \label{fig:Si10H16}
     \end{subfigure}
     \hfill
     \begin{subfigure}[b]{0.4\linewidth}
         \centering
         \includegraphics[width=\linewidth]{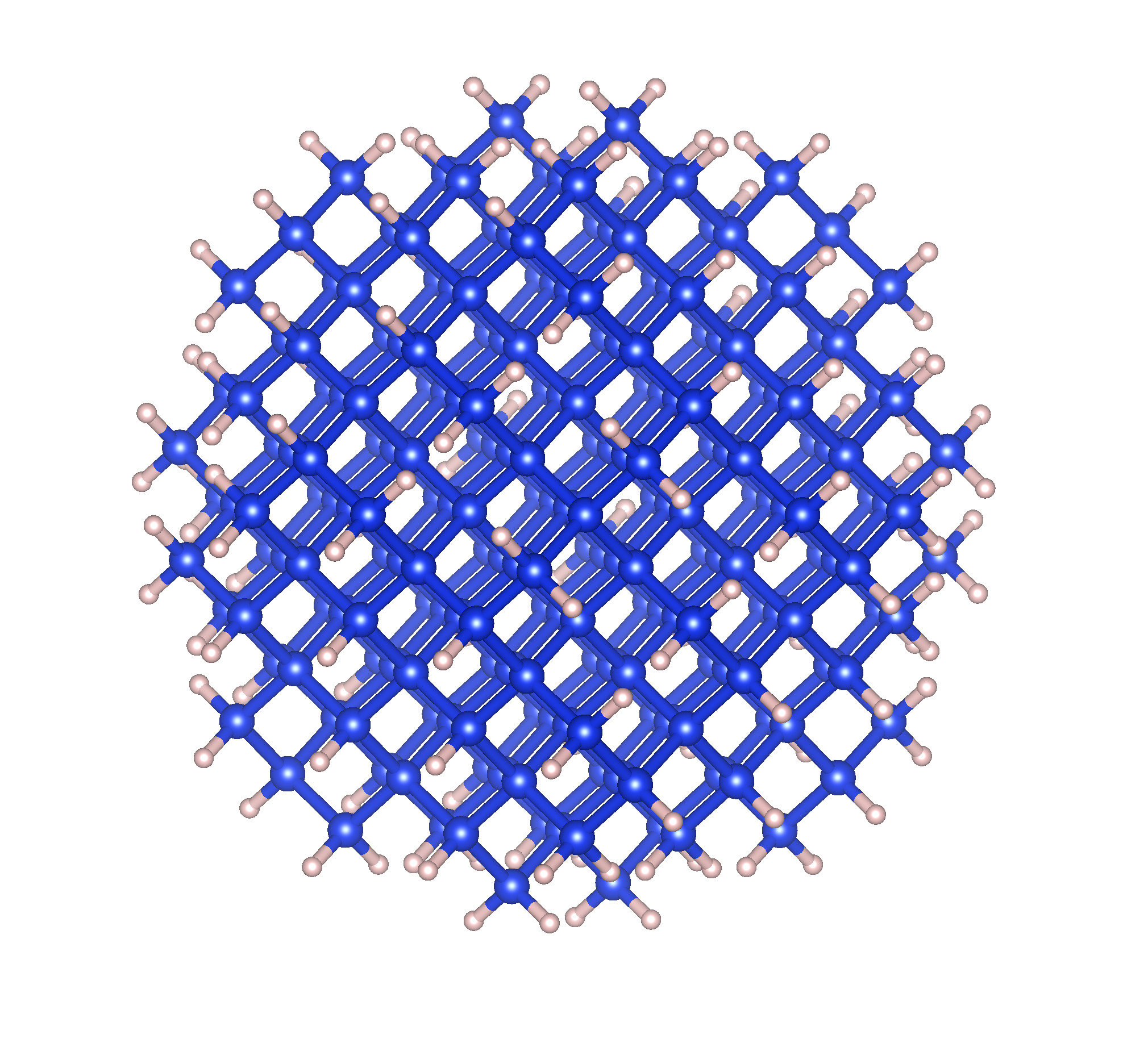}
         \caption{$\mathrm{Si}_{220}\mathrm{H}_{144}$}
         \label{fig:Si220H144}
     \end{subfigure}
     \caption{Schematics of the benchmark silicon quantum dots. 
     }
     \label{fig:Si bns model}
\end{figure}

In the present work, besides providing the code for the TTDFT calculation, we focus on optimizing the most computationally expensive part of the calculation---the repetitive matrix-matrix multiplication kernel called during the Chebyshev filtering step---and further using GPU acceleration to improve the computational efficiency of the calculations. In order to assess the optimization realized, we use the benchmark systems from our previous work~\cite{Lin2021} comprising of aluminum nano-particles and silicon quantum dots of various sizes. The aluminum nano-particles ranging from $\mathrm{Al}_{13}$ to $\mathrm{Al}_{6525}$ are constructed using icosahedral symmetry. The silicon quantum dots are constructed by rounding the diamond-structured silicon crystal and passivating the surface with hydrogen atoms. The silicon quantum-dots considered here range from $\mathrm{Si}_{10}\mathrm{H}_{16}$ to $\mathrm{Si}_{6047}\mathrm{H}_{1308}$. Ball and stick models for the two smallest clusters of both systems are depicted in Fig.~\ref{fig:Al bns model} and Fig.~\ref{fig:Si bns model}. In order to conduct a performance analysis, we ran the benchmark calculations by solely using CPUs and compared with the acceleration obtained by utilizing GPUs for the matrix-matrix multiplication kernel in the Chebyshev filtering and subspace projection steps. Further, to ensure the accuracy of the code with GPU acceleration, we compare the ground-state energies obtained via CPU-only and CPU-GPU calculations of the two smallest clusters for both systems.

The numerical parameters used in the present study follow the previously converged CPU-based calculations of the benchmark systems~\cite{Lin2021}. In this work, we use local density approximation (LDA) for exchange-correlation functional~\cite{LDA_CA, LDA_PW, LDA_PZ} and a norm-conserving Troullier-Martin pseudopotential in Kleinmann-Bylander form~\cite{PSP_TM, PSP_KB}. The Fermi-Dirac smearing temperature is set to $T=500K$ for computing the fractional occupancy of the orbitals. The Chebyshev polynomial degree is chosen to be $10-20$ for various materials systems. The Tucker decomposition ranks~\cite{Lin2021} in the evaluation of the Hartree potential ($R_{H}$), in the representation of local part of the effective Kohn-Sham potential ($R_{V}$) and the non-local part of the effective potential ($R^{nl}_{V}$) are chosen to be $R_{H}=40$, $R_{V}=50$, $R^{nl}_{V}=25$ for all aluminum nano-particles system and $R_{H}=55$, $R_{V}=55$ and $R^{nl}_{V}=25$ for all silicon quantum dots system. The prescribed truncation tolerance for the Kohn-Sham Hamiltonian is set to $10^{-4}$ Ha for both aluminum nano-particles and silicon quantum dots according to the previous error analysis in~\cite{Lin2021}. The numerical parameters used are consistent for both the CPU- and GPU-based calculations in the performance analysis.  

The performance benchmarks have been conducted on the Summit supercomputer, with each node comprising of 2 IBM Power 9 CPUs (with 42 physical cores) and 6 NVIDIA Tesla V100 GPUs.  We note that the number of nodes used to conduct the calculations is  chosen such that the calculation is within the good parallel-scaling regime to obtain a representative measure of computational efficiency. In particular, for the larger systems considered in this work, the number of nodes are chosen such that the number of rows of $\mathbf{H}_L$---the Kohn-Sham Hamiltonian matrix in the localized Tucker tensor basis---owned by each GPU is around 30,000, which maintains a good balance between memory limitation and parallel scaling efficiency.

\subsection{Accuracy analysis of CPU and GPU implementation}
In this section, we perform full ground-state calculations for the smaller benchmark systems of both aluminum nano-particles and silicon quantum dots to verify that the GPU implementation provides the same results as the CPU implementation. For the aluminum nano-particles, we choose $\mathrm{Al}_{13}$ and $\mathrm{Al}_{147}$. The ball and stick model for the systems are provided in Fig.~\ref{fig:Al bns model}. For the silicon quantum dots, $\mathrm{Si}_{10}\mathrm{H}_{16}$ and $\mathrm{Si}_{220}\mathrm{H}_{144}$ are investigated. The ball and stick model for the silicon quantum dots are illustrated in Fig.~\ref{fig:Si bns model}. The converged results are tabulated in Table~\ref{tab:accuracy}. We note that the results from the CPU-based and GPU-based calculations are identical up to the chemical accuracy of interest, with the differences at $\mathcal{O}(10^{-8})$ eV in the ground-state energy per atom. This small difference is possibly a result of the round-off error accumulations during the course of the ground-state calculation. 

\begin{table*}[h!]
\centering
\begin{tabular}{|c|c|c|c|c|}
\hline
 & $\mathrm{Al_{13}}$ & $\mathrm{Al_{147}}$ & $\mathrm{Si_{10}H_{16}}$ & $\mathrm{Si_{220}H_{144}}$ \\ \hline
CPU & -55.996571 & -56.617932 & -51.027192 & -71.384192 \\ \hline
GPU & -55.996571 & -56.617932 & -51.027192 & -71.384192 \\ \hline
$\abs{\mathrm{Error}}$ & $8.97 \times 10^{-9}$ & $4.35 \times 10^{-8}$ & $9.26 \times 10^{-9}$ & $1.21 \times 10^{-8}$ \\ \hline
\end{tabular}
\caption{Accuracy comparison of CPU and GPU implementation in ground state energy per atom (eV) for $\mathrm{Al_{13}}$, $\mathrm{Al_{147}}$, $\mathrm{Si_{10}H_{16}}$, and $\mathrm{Si_{220}H_{144}}$.}
\label{tab:accuracy}
\end{table*}

\subsection{Performance analysis} \label{ch:performance analysis for aluminum nano-particles}
In order to assess the computational efficiency derived from GPU acceleration, we compare the single SCF execution time in node-hours for the CPU-based implementation with that of the GPU acceleration. The computational times (in node-hours) for the various benchmark systems are provided in Table~\ref{tab:Al time breakdown} and Table~\ref{tab:Si time breakdown} for the aluminum nano-particles and silicon quantum dots, respectively. In particular, the breakdown of the single SCF computational time is provided for all the major steps of Algorithm~1: (i) ChF: Chebyshev filtering step; (ii) Orth: Orthogonalization of the Chebyshev filtered vectors; (iii) Sub proj: Projection of the Kohn-Sham Hamiltonian matrix onto the Chebyshev filtered subspace; (iv) Others: all other costs in the SCF, including solution of the eigenvalue problem in the Chebyshev filtered subspace. The total computational cost for a single SCF iteration and the speedup from GPU acceleration is also provided.  

\subsubsection{Aluminum nano-particles}
Table~\ref{tab:Al time breakdown} shows the single SCF breakdown of the computational times of the CPU-based calculations and that of the GPU accelerated calculations for aluminum nano-particles of various sizes. We remark that the main focus of this work is to optimize the matrix-matrix multiplication kernel in the Chebyshev filtering step---the most expensive step in each SCF iteration---using GPU acceleration with the approach proposed in Sec.~\ref{ch:HX implementation}. Thus, the orthogonalization of the Chebyshev filtered vectors and other parts of in the calculations are essentially done on CPUs, which are identified using * in the table. For the smallest benchmark system considered ($\mathrm{Al}_{13}$), we note that the calculation using GPU acceleration is slower than the CPU-based calculation, and the performance is comparable for $\mathrm{Al}_{147}$. This is due to the overhead costs for transferring data  between the host and  the device memory that are competing in small system sizes with the arithmetic efficiency gained from GPU acceleration. However, for all the other systems, we obtain overall GPU acceleration in the Chebyshev filtering step and the subspace projection step---the two parts of the algorithm that are affected by the GPU acceleration of the matrix-matrix multiplication kernel. In particular, for the largest system size considered, $\mathrm{Al}_{6525}$ nano-particle, we obtain $\sim7.8\times$ and $\sim 8.2 \times$ computational efficiency in the Chebyshev filtering step and the subspace projection step, respectively, due to the GPU acceleration. This, in turn, provides a $\sim 3.1\times$ improvement in the computational efficiency for the SCF iteration step that is representative of the full ground-state calculation. We note that with possible GPU acceleration of the orthogonalization step, the potential improvement in the computational efficiency by using GPU acceleration can even be greater.

\begin{table*}[h!]
\centering
\begin{tabular}{|c|c|c|c|c|c|c|c|}
\hline
 &
   &
  \begin{tabular}[c]{@{}c@{}}ChF\\ (node-hrs)\end{tabular} &
  \begin{tabular}[c]{@{}c@{}}Orth*\\ (node-hrs)\end{tabular} &
  \begin{tabular}[c]{@{}c@{}}Sub proj\\ (node-hrs)\end{tabular} &
  \begin{tabular}[c]{@{}c@{}}Others*\\ (node-hrs)\end{tabular} &
  \begin{tabular}[c]{@{}c@{}}Time/SCF\\ (node-hrs)\end{tabular} &
  $\mathrm{\frac{GPU}{CPU}}$ \\ \hline 
\multirow{2}{*}{$\mathrm{Al_{13}}$}   & CPU & 1.87E-04 & 8.70E-05 & 3.57E-05 & 1.42E-04 & 4.52E-04 & \multirow{2}{*}{0.85} \\ \cline{2-7}
                                      & GPU & 2.34E-04 & 7.24E-05 & 3.12E-05 & 1.92E-04 & 5.30E-04 &                       \\ \hline
\multirow{2}{*}{$\mathrm{Al_{147}}$}  & CPU & 3.21E-02 & 2.14E-04 & 5.32E-03 & 1.12E-02 & 4.88E-02 & \multirow{2}{*}{1.39} \\ \cline{2-7}
                                      & GPU & 2.19E-02 & 2.02E-04 & 3.21E-03 & 9.80E-03 & 3.51E-02 &                       \\ \hline
\multirow{2}{*}{$\mathrm{Al_{561}}$}  & CPU & 0.331    & 0.008    & 0.047    & 0.156    & 0.542    & \multirow{2}{*}{1.75} \\ \cline{2-7}
                                      & GPU & 0.135    & 0.009    & 0.013    & 0.152    & 0.309    &                       \\ \hline
\multirow{2}{*}{$\mathrm{Al_{2057}}$} & CPU & 3.724    & 0.228    & 0.482    & 1.327    & 5.761    & \multirow{2}{*}{2.97} \\ \cline{2-7}
                                      & GPU & 0.513    & 0.212    & 0.099    & 1.119    & 1.943    &                       \\ \hline
\multirow{2}{*}{$\mathrm{Al_{6525}}$} & CPU & 30.119   & 6.192    & 3.422    & 4.132    & 43.865   & \multirow{2}{*}{3.12} \\ \cline{2-7}
                                      & GPU & 3.872    & 5.871    & 0.415    & 3.891    & 14.049   &                       \\ \hline
\end{tabular}%
\caption{Breakdown of single-SCF computational times (in node-hours) for CPU-based and GPU-based calculations for the benchmark systems comprising of Al nano-particles. The columns marked with asterisk * are computed on the host (CPU) without GPU optimization.}
\label{tab:Al time breakdown}
\end{table*}

\subsubsection{Silicon quantum dots} \label{ch:performance analysis for silicon quantum dots}
Table~\ref{tab:Si time breakdown} shows the single SCF breakdown of computational times of CPU-based and GPU-accelerated calculations for various sizes of silicon quantum dots. Similar to the aluminum nano-particles, the benefits of GPU acceleration improve with system size. Notably, for the largest quantum dot system $\mathrm{Si}_{6047}\mathrm{H}_{1308}$ which contains 6355 atoms, the computational efficiency gain by using GPU acceleration is $\sim 7.2\times$ in Chebyshev filtering step and $\sim 6.8\times$ for the subspace projection step. The computational efficiency gain for the full SCF iteration is $\sim 3.4\times$.

\begin{table*}[h!]
\centering
\begin{tabular}{|c|c|c|c|c|c|c|c|}
\hline
 &
   &
  \begin{tabular}[c]{@{}c@{}}ChF\\ (node-hrs)\end{tabular} &
  \begin{tabular}[c]{@{}c@{}}Orth*\\ (node-hrs)\end{tabular} &
  \begin{tabular}[c]{@{}c@{}}Sub proj\\ (node-hrs)\end{tabular} &
  \begin{tabular}[c]{@{}c@{}}Others*\\ (node-hrs)\end{tabular} &
  \begin{tabular}[c]{@{}c@{}}Time/SCF\\ (node-hrs)\end{tabular} &
  $\mathrm{\frac{GPU}{CPU}}$ \\ \hline
\multirow{2}{*}{$\mathrm{Si_{10}H_{16}}$}     & CPU & 1.37E-03 & 5.99E-04 & 1.83E-04 & 2.42E-03 & 4.57E-03 & \multirow{2}{*}{1.02} \\ \cline{2-7}
                                              & GPU & 1.52E-03 & 5.87E-04 & 1.79E-04 & 2.21E-03 & 4.50E-03 &                       \\ \hline
\multirow{2}{*}{$\mathrm{Si_{220}H_{144}}$}   & CPU & 6.12E-02 & 1.10E-03 & 6.99E-03 & 2.33E-02 & 9.26E-02 & \multirow{2}{*}{1.72} \\ \cline{2-7}
                                              & GPU & 2.79E-02 & 1.21E-03 & 4.62E-03 & 2.01E-02 & 5.38E-02 &                       \\ \hline
\multirow{2}{*}{$\mathrm{Si_{525}H_{276}}$}   & CPU & 0.515    & 0.018    & 0.067    & 0.214    & 0.814    & \multirow{2}{*}{1.9}  \\ \cline{2-7}
                                              & GPU & 0.203    & 0.017    & 0.028    & 0.180    & 0.428    &                       \\ \hline
\multirow{2}{*}{$\mathrm{Si_{1214}H_{504}}$}  & CPU & 2.132    & 0.132    & 0.258    & 0.552    & 3.074    & \multirow{2}{*}{2.47} \\ \cline{2-7}
                                              & GPU & 0.611    & 0.127    & 0.087    & 0.422    & 1.247    &                       \\ \hline
\multirow{2}{*}{$\mathrm{Si_{6047}H_{1308}}$} & CPU & 38.511   & 6.525    & 4.259    & 3.515    & 52.810   & \multirow{2}{*}{3.36} \\ \cline{2-7}
                                              & GPU & 5.385    & 6.473    & 0.629    & 3.223    & 15.710   &                       \\ \hline
\end{tabular}%
\caption{Breakdown of single-SCF computational times (in node-hours) for CPU-based and GPU-based calculations for the benchmark systems comprising of silicon quantum dots. The columns marked with asterisk * are computed on the host (CPU) without GPU optimization. 
}
\label{tab:Si time breakdown}
\end{table*}

\section{Summary} \label{ch:summary}
We have presented the TTDFT code with GPU acceleration for the main compute intensive kernels of the calculation. In particular, the TTDFT algorithm is based on using a systematically convergent localized basis that is generated from an additive separable approximation of the Kohn-Sham Hamiltonian~\cite{Lin2021}. The solution to the discrete Kohn-Sham problem is computed via Chebyshev filtering subspace iteration~\cite{Zhou2006} method. The compute intensive kernels in the TTDFT code that involve matrix-matrix multiplication of a symmetric sparse matrix (Hamiltonian matrix) and a dense matrix (wavefunction matrix) have been GPU accelerated. The benchmark studies show a substantial improvement for the GPU-accelerated steps of the algorithm---\,$\sim8\times$ for the largest system sizes---which improves the overall computational efficiency of the calculation. We note that recent studies have shown that the TTDFT algorithm can substantially outperform plane-wave implementations for large-scale systems~\cite{Lin2021}, owing to the reduced-order scaling with system size. The present GPU-based TTDFT code is a further step towards enabling systematically convergent and computationally efficient large-scale DFT calculations. 

We note that upon GPU accelerating the compute intensive kernels, the main computational bottleneck in the TTDFT code has now shifted to the step involving orthogonalization of the Chebyshev filtered vectors. Developing an efficient GPU implementation of the orthogonalization procedure and GPU porting other parts of the code can further improve the performance of the TTDFT code, and is a useful direction to pursue.

\section*{Acknowledgments}
\noindent We gratefully acknowledge the support of the Air Force Office of Scientific Research through grant number FA9550-21-1-0302 under the auspices of which this work was conducted. V.G. also gratefully acknowledges the support of the Army Research Office through the DURIP grant W911NF1810242, which provided computational resources for this work. This research also used resources of the Oak Ridge Leadership Computing Facility, which is a DOE Office of Science User Facility supported under Contract DE-AC05-00OR22725.




\bibliographystyle{elsarticle-num}



\end{document}